\documentclass[final,twocolumn,10pt,titlepage,prb]{revtex4}%
\usepackage{amsfonts}
\usepackage{amsmath}
\usepackage{amssymb}
\usepackage{graphicx}%
\providecommand{\U}[1]{\protect\rule{.1in}{.1in}}
\begin{document}
\preprint{ }
\title{Magnetic interactions of substitutional Mn pairs in GaAs}
\author{T. O. Strandberg}
\affiliation{School of Pure and Applied Natural Sciences, Kalmar University, 391 82 Kalmar, Sweden}
\author{C. M. Canali}
\affiliation{School of Pure and Applied Natural Sciences, Kalmar University, 391 82 Kalmar, Sweden}
\author{A. H. MacDonald}
\affiliation{Department of Physics, University of Texas at Austin, Austin, Texas 78712, USA}

\begin{abstract}
We employ a kinetic-exchange tight-binding model to calculate the magnetic
interaction and anisotropy energies of a pair of substitutional Mn atoms in
GaAs as a function of their separation distance and direction. We find that
the most energetically stable configuration is usually one in which the spins
are ferromagnetically aligned along the vector connecting the Mn atoms. The
ferromagnetic configuration is characterized by a splitting of the topmost
unoccupied acceptor levels, which is visible in scanning tunneling microscope
studies when the pair is close to the surface and is strongly dependent on
pair orientation. The largest acceptor splittings occur when the Mn pair is
oriented along the $\langle110\rangle$ symmetry direction, and the smallest
when they are oriented along $\langle100\rangle$. We show explicitly that the
acceptor splitting is not simply related to the effective exchange interaction
between the Mn local moments. The exchange interaction constant is instead
more directly related to the width of the distribution of all impurity levels
-- occupied and unoccupied. When the Mn pair is at the (110) GaAs surface,
both acceptor splitting and effective exchange interaction are very small
except for the smallest possible Mn separation.

\end{abstract}

\maketitle

\section{Introduction}

Experimental progress in the past 5 years has led to a large number of
experimental studies of magnetic and nonmagnetic transition-metal impurities
in semiconductors using advanced scanning tunneling microscope (STM)
techniques.\cite{yakunin_prl04,
yakunin_prl05,kitchen_jsc05,yazdani_nat06,yazdani_jap07,koenraad_prb08,yazdani_prb09}
This effort was motivated in part by the hope that the high-resolution imaging
and spatially resolved spectroscopic power of the STM could help in developing
an accurate microscopic picture of dilute magnetic semiconductor (DMS)
magnetism. In DMSs magnetic impurities provide local moments, which can couple
to yield a collective ferromagnetic state. In the prototypical DMS, (Ga,Mn)As,
the Mn impurities act as acceptors providing itinerant holes that can mediate
long-range interactions between local moments. Among the open issues in DMS
physics\cite{dietl_spintronics_08} are the precise character of the hole
states, the nature of the coupling of the holes with the local magnetic
moments, and the properties of the ensuing magnetic interaction between local
moments. STM experiments performed
recently\cite{yazdani_nat06,yazdani_jap07,koenraad_prb08,yazdani_prb09} are
playing a decisive role in clarifying some of these issues.

In Ref.~[\onlinecite{yazdani_nat06}] STM substitution techniques were used to
incorporate individual Mn atoms into Ga sites in a GaAs (110) surface.
Real-space spectroscopic measurements in the vicinity of an isolated Mn
impurity revealed the presence of a mid-gap resonance arising from a Mn
induced acceptor state. High-resolution imaging showed that the acceptor wave
function is strongly anisotropic with respect to the crystal axes of the host.
When two Mn atoms were incorporated close to each other, two resonances
appeared in the gap, split by approximately 0.5 eV. The splitting was found to
be strongly dependent on the Mn pair orientation with respect to the GaAs
crystal structure and on Mn separation. A simple toy model, describing
acceptor states coupled to the Mn ion local moments of the two
impurities,\cite{schilf_prb01} suggested that a measurable splitting of the
acceptor levels could only occur if the two Mn local moments were
ferromagnetically aligned.\cite{yazdani_nat06} Since acceptor-level splitting
is an observable indicator of ferromagnetic coupling, it seemed plausible that
the dependence of this splitting on separation should be related at least
qualitatively to the Mn-Mn exchange interaction. If so, the STM experiment
could be used to measure exchange interactions between the Mn moments and test
theories of this interaction. One of the purposes of the present study is to
examine this relationship quantitatively.

Several of the experimental features uncovered in
Ref.~[\onlinecite{yazdani_nat06}] could be qualitatively accounted for
theoretically by a tight-binding model calculation for Mn in bulk GaAs
presented in the same paper. A more thorough comparison between experiments
and theoretical modeling, both for Mn atoms in the bulk \textit{and Mn near
the surface} is nevertheless necessary to interpret the experiments,
motivating the present theoretical work. Here we consider a kinetic $p$-$d$
exchange tight-binding model\cite{scm_MnGaAs_paper1_prb09} in which the
effective exchange interaction between the hole states and the local Mn
moments arises from hybridization of the impurity $d$ levels with $p$ levels
of the host. The model is solved numerically for large super-clusters
containing up to 3200 atoms. This approach allows us to place the Mn pair
either in bulk GaAs or on the (110) surface.

Within this model, we study the electronic and magnetic properties of Mn pairs
in GaAs, assuming that the two local magnetic moments are collinear, having
either parallel [ferromagnetic (FM)] or antiparallel [antiferromagnetic (AFM)]
relative orientation. One of our goals is to study the spin-orbit induced
magnetic anisotropy energy of the system and see how this quantity is related
to the properties of the mid-gap acceptor states. In the FM configuration the
magnetic moment tends to point along the direction of the pair, while in the
AFM configuration there is typically a quasi-easy plane perpendicular to the
pair direction. As in the isolated Mn impurity case previously
studied,\cite{scm_MnGaAs_paper1_prb09} we find that the sum of the individual
anisotropy energies of the top two unoccupied valence band orbitals, mirrors
the total anisotropy of the system. This shows that the picture of Mn-Mn
interactions mediated by valence band holes is valid, and simplifies the
interpretation of our numerical results. We then consider the properties of
the acceptor levels for the two possible relative orientations of the Mn
magnetic moments. In the FM configuration, which is generally found to be the
ground state for most Mn pair orientations and Mn separations, the acceptor
levels lie in a group just above the valence-band maximum and have a splitting
that is enhanced by inter-ion hybridization. The group of six split levels can
be viewed as a nascent version of the impurity band which forms in the bulk at
small but finite Mn concentrations.
\footnote{\label{foot1}In the following, we will make a loose use of the
expression \textit{impurity band} to indicate this group of six split impurity
levels in the gap.} In particular, the two unoccupied acceptor levels (i.e.
occupied by holes), which fully determine both magnetic interaction and
anisotropy energies have a finite splitting, which can be measured in
experiment. For Mn pairs in bulk, we find that the acceptor splitting varies
strongly with pair orientation and Mn separation, in agreement with experiment
and previous calculations.\cite{yazdani_nat06} The splitting is maximal for
the most closely spaced Mn pair oriented along the $\langle110\rangle$
direction, where it is of the order of few hundred meV, and very small for
pairs oriented along $\langle100\rangle$. For some Mn pairs, the wave
functions of the two acceptor states have bonding and antibonding character,
which is again consistent with experiment.\cite{yazdani_nat06, yazdani_jap07}
These results support the validity of the kinetic $p$-$d$ exchange model.

In our study the energy difference between FM and AFM moment orientations is
related to partial occupation of the acceptor impurity levels, which are split
more widely in the FM configuration. When we calculate the exchange constant
$J$ for the Mn-Mn interaction, as the difference between the ground-state
energies of the two configurations, we find that $J$ is not in one-to-one
correspondence with the splitting between the two unoccupied acceptor levels.
In particular, the level splitting is much more anisotropic than $J$ with
respect to the Mn pair orientation. On the other hand, we find that estimates
of the width of the distribution of all impurity levels -- occupied and
unoccupied -- are more directly related to $J$.

Effective exchange interactions between Mn ions in bulk (Ga,Mn)As have been
calculated by several groups using either \emph{ab initio} methods or more
phenomenological approaches. Comparison of our results with other estimates is
not always straightforward, since we have only two Mn moments and the exchange
interactions are not strictly pairwise. The order of magnitude of our $J$ is
nevertheless consistent with published results obtained from first-principles
calculations, although values of $J$ for specific Mn pair directions and
separations may differ. However, it is well-known that DMSs are not accurately
described by the local-density-approximation often used in first principles
methods. The discrepancies could either be due to the shortcomings of our
model or to the inaccuracies of \emph{ab initio} calculations.

We have also looked at how the Mn pairs interact when they are placed in a
(110) GaAs surface. Typically, we find that the acceptor wave functions become
highly localized at the surface and produce states that are deep in the band
gap. As a result, the long-range interactions are much weaker and
antiferromagnetic alignment of Mn spins is more likely to occur. The magnetic
anisotropy energies at the surface are an order of magnitude smaller than in
bulk, and tend to produce quasi-easy planes at close distances, reverting to
isolated Mn anisotropy landscapes at larger separations. For Mn pairs in the
(110) surface, the long-range behavior of the acceptor wave functions and
acceptor splitting seems to agree less well with experiment than for Mn pairs
in bulk GaAs. The tight-binding model that we use is of course less well
justified when the Mn pair is in the surface. However, effects present in the
experiment may contribute to this difference. In addition to an uncertainty in
model parameters at the surface due to band-bending shifts of the acceptors, a
large overlap with continuum states due to Zn co-dopants in the sample can
cause a more extended, bulk-like Mn acceptor wave function. Beyond the scope
of the present paper, care should also be taken in simulating the change in
the effective potential experienced by surface electrons due to the addition
and removal process introduced by the STM. Our results clearly demonstrate
that modeling STM studies of the surface of a system as complicated as
(Ga,Mn)As is highly nontrivial and requires additional work.

The paper is organized as follows. In Sec.~\ref{theorysection} we review some
theoretical aspects of the (Ga,Mn)As system and give a brief introduction to
our tight-binding Hamiltonian. We also elaborate on a toy model that gives an
idea of the system behavior expected when the Mn moments are parallel or
antiparallel. The results for pairs of Mn along different symmetry directions
in a fully periodic bulk-like environment are presented in
Sec.~\ref{bulkresults}, and in Sec.~\ref{surfresults} for Mn pairs in the
(110) GaAs surface. Finally, we summarize our conclusions in
Sec.~\ref{conclusions}.

\section{Theory}

\label{theorysection}

\begin{figure}[ptb]
\resizebox{7.5cm}{!}{\includegraphics{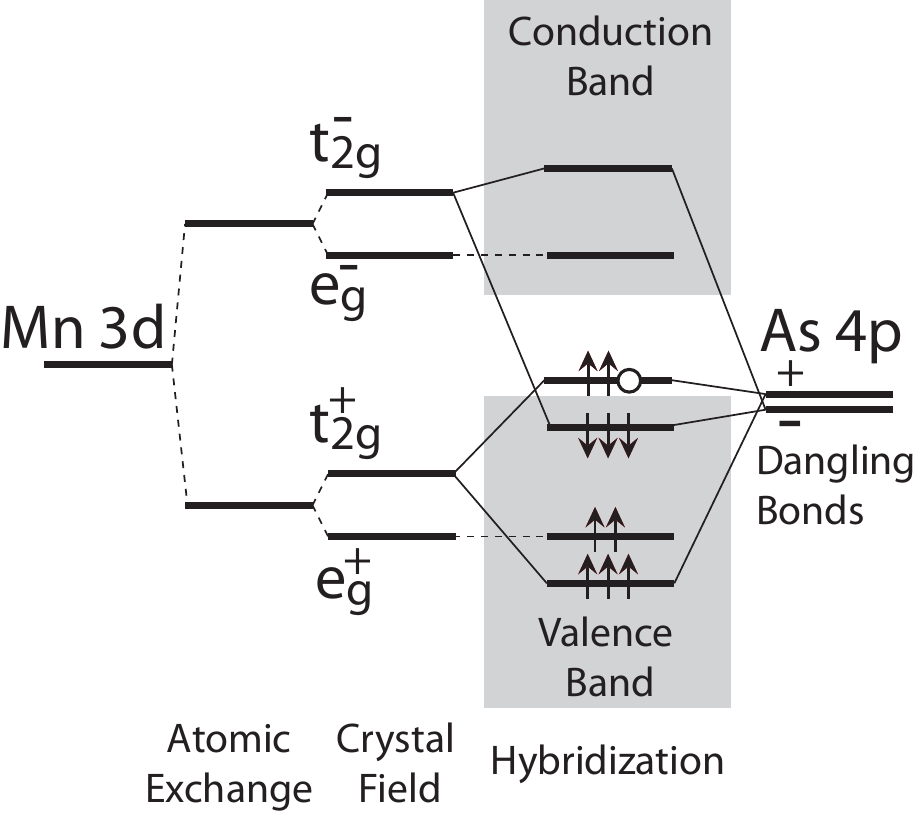}}\caption{Schematic drawing
of the level dynamics leading to a polarized acceptor at the top of the
valence band. The Mn $3d^{5}$ up and down electrons are split by the atomic
exchange in accordance with Hund's first rule. The levels are further split by
the crystal field into a doublet of $e_{g}$ and a triplet of $t_{2g}%
$-symmetry. The $t_{2g}$-levels hybridize with nearest neighbor dangling bond
As $p$-states, such that the levels at the top of the valence band with spin
parallel to the Mn spin are pushed up in energy. Spin up is denoted by + and
spin down by -.}%
\label{cartoon2}%
\end{figure}

\subsection{Hamiltonian for Mn impurities in GaAs}

\label{Hamiltonian_MnGaAs} In this section we briefly review the basic physics
of Mn impurities in GaAs, which motivates the choice of the tight-binding
Hamiltonian\cite{scm_MnGaAs_paper1_prb09} used throughout the paper. For more
details the reader is referred to Ref.~[\onlinecite{scm_MnGaAs_paper1_prb09}],
where the same model was used to investigate the properties of individual Mn
atoms in GaAs.

In the neutral state, consisting of the Mn$^{2+}$ ion on a Ga site with a
weakly bound hole, the Mn has spin $S=5/2$ and orbital moment $L=0$%
.\cite{jungw_rmp06} This is in accordance with Hund's first rule that survives
even when the Mn is embedded in the crystal. The Mn $3d^{5}$ electrons build
up large, localized magnetic moments. Because the Mn is missing the Ga 4$p$
valence electron, it introduces a hole in the system, simultaneously acting as
an acceptor and a source of magnetic moments. In bulk, the holes introduced by
many Mn impurities form itinerant carriers that mediate the ferromagnetic
coupling between localized moments. Fig.~\ref{cartoon2} shows a schematic
drawing of what happens when a Mn substitutes a Ga in the GaAs
lattice.\cite{zhao_apl04} The Mn $3d^{5}$ up states (for example) that lodge
in the valence band are exchange-split from the down states that end up in the
conduction band. The crystal field imposes the tetrahedral host symmetry,
resulting in a further split into a doublet of $e_{g}$ symmetry and a triplet
of $t_{2g}$ symmetry.\cite{vogl_app05} The $e_{g}$ states couple only weakly
to the host. The $t_{2g}$ states on the other hand, hybridize with the As
nearest neighbor dangling-bond $p$ states at the top of the valence band,
forming bonding and antibonding combinations. Effectively, the neighboring As
$p$ spins at the top of the valence band that are parallel to the Mn spin are
shifted up in energy, whereas $p$ spins that are antiparallel to the Mn spin
are shifted down. Therefore, the main effect of a single substitutional Mn
with spin up, is to introduce three spin-polarized ("up") levels above the top
of the valence band and three spin-polarized ("down") levels below. These six
states are predominantly of As $p$ character. The hole introduced by a single
Mn will occupy the highest of the three antibonding states (indicated by the
empty circle in Fig.~\ref{cartoon2}). In this way, the hybridization mechanism
gives rise to an antiferromagnetic coupling between itinerant and localized
spins by what is known as Zener's kinetic-exchange
mechanism.\cite{zener_pr51,dietl_sci00,bhattacharjee_pbc83,okabayashi_prb98}
We account for this $p$-$d$ indirect exchange by introducing a classical Mn
vector of magnitude 5/2, which couples to the neighboring As $p$ states. There
are no explicit $d$ electrons in our model; they are accounted for only
implicitly via the exchange interaction between the localized moments and the
As $p$ spins on neighboring sites.

Our tight-binding kinetic exchange Hamiltonian takes the following form:%
\begin{align}
H  &  =\sum_{ij,\mu\mu^{\prime},\sigma}t_{\mu\mu^{\prime}}^{ij}a_{i\mu\sigma
}^{\dag}a_{j\mu^{\prime}\sigma}^{\phantom{\dag}}+J_{pd}\sum_{m}\sum_{n[m]}%
\vec{S}_{n}\cdot\hat{\Omega}_{m}\nonumber\\
&  +\sum_{i,\mu\mu^{\prime},\sigma\sigma^{\prime}}\lambda_{i}\langle\mu
,\sigma|\vec{L}\cdot\vec{S}|\mu^{\prime},\sigma^{\prime}\rangle a_{i\mu\sigma
}^{\dag}a_{i\mu^{\prime}\sigma^{\prime}}^{\phantom{\dag}}\nonumber\\
&  +\frac{e^{2}}{4\pi\varepsilon_{0}\varepsilon_{r}}\sum_{m}\sum_{i\mu\sigma
}\frac{a_{i\mu\sigma}^{\dag}a_{i\mu\sigma}^{\phantom{\dag}}}{|\vec{r}%
_{i}\mathbf{-}\vec{R}_{m}|}+V_{\mathrm{corr}}. \label{hamiltonian}%
\end{align}
The first term reproduces the band-structure of bulk GaAs\cite{chadi_prb77}
and contains the near-neighbor hopping and on-site energies in terms of the
Slater-Koster parameters\cite{slaterkoster_pr54,papac_jpcm03} $t_{\mu
\mu^{\prime}}^{ij}$ for the $s$ and $p$ electrons of Ga and As. In
(\ref{hamiltonian}) $i$ and $j$ are atomic indices, $\mu$ and $\mu^{\prime}$
are orbital indices and $\sigma$ denotes spin. In simulating the (110)
surface, buckling is accounted for by rescaling\cite{chadi_prl78,chadi_prb79}
the tight-binding parameters and modifying the direction cosines appropriately.

The second term implements the $p$-$d$ exchange mechanism described above. It
couples the unit spin vector of Mn atom $m$: $\hat{\Omega}_{m}$, to the $p $
orbitals of the nearest neighbor As atoms $n[m]$: $\vec{S}_{n}=\tfrac{1}%
{2}~\sum_{\pi\sigma\sigma^{\prime}}a_{n\pi\sigma}^{\dag}\vec{\tau}%
_{\sigma\sigma^{\prime}}a_{n\pi\sigma^{\prime}}$, where $\vec{\tau}$ is a
vector of Pauli matrices. The value of $J_{pd}$ has been inferred by
theory\cite{timmacd_prb05} and experiment\cite{ohno_sci98} to approximately
$J_{pd}=1.5$ eV. The three $p$-$d$ hybridized levels, spin-polarized in the
direction of the Mn moment $\hat{\Omega}_{m}$ are split from the three levels
polarized in the opposite direction by an energy of order $J_{pd}$.

The third term in Eq.~(\ref{hamiltonian}) accounts for SO interactions in an
atomic approximation with the renormalized spin-orbit
splittings\cite{chadi_prb77} $\lambda_{\text{Ga}}=2\lambda_{\text{Mn}}=58$ meV
and $\lambda_{\text{As}}=140$ meV. Spin-orbit interaction causes the band
energy to depend not only on the relative angles between different Mn spin
directions, but also on spin-orientations relative to the lattice.

The fourth term of the Hamiltonian accounts for a long-range repulsive Coulomb
interaction in the presence of a Mn impurity, which attracts a weakly bound
hole and repels electrons. At the surface, we crudely account for the weaker
dielectric screening by reducing the bulk dielectric constant $\varepsilon
_{r}$ in half. The Coulomb behavior at short distances is parametrized in
$V_{\mathrm{corr}},$ which is a Mn central cell correction term used as a
parameter to tune the Mn acceptor level to the experimental
position.\cite{schairer_prb74,lee_ssc64,chapman_prl67,linnarsson_prb97} It
contains an on-site $V_{\mathrm{on}}=1.0$ eV (that is, acting on the Mn) and
an off-site term $V_{\mathrm{off}}=2.4$ eV. The off-site Coulomb correction
affects the nearest neighbor As atoms and together with $J_{pd}$ controls the
amount of $p$-$d$ hybridization in the system and the range of the acceptor
wave function.

\begin{figure}[ptb]
\resizebox{5.3cm}{!}{\includegraphics{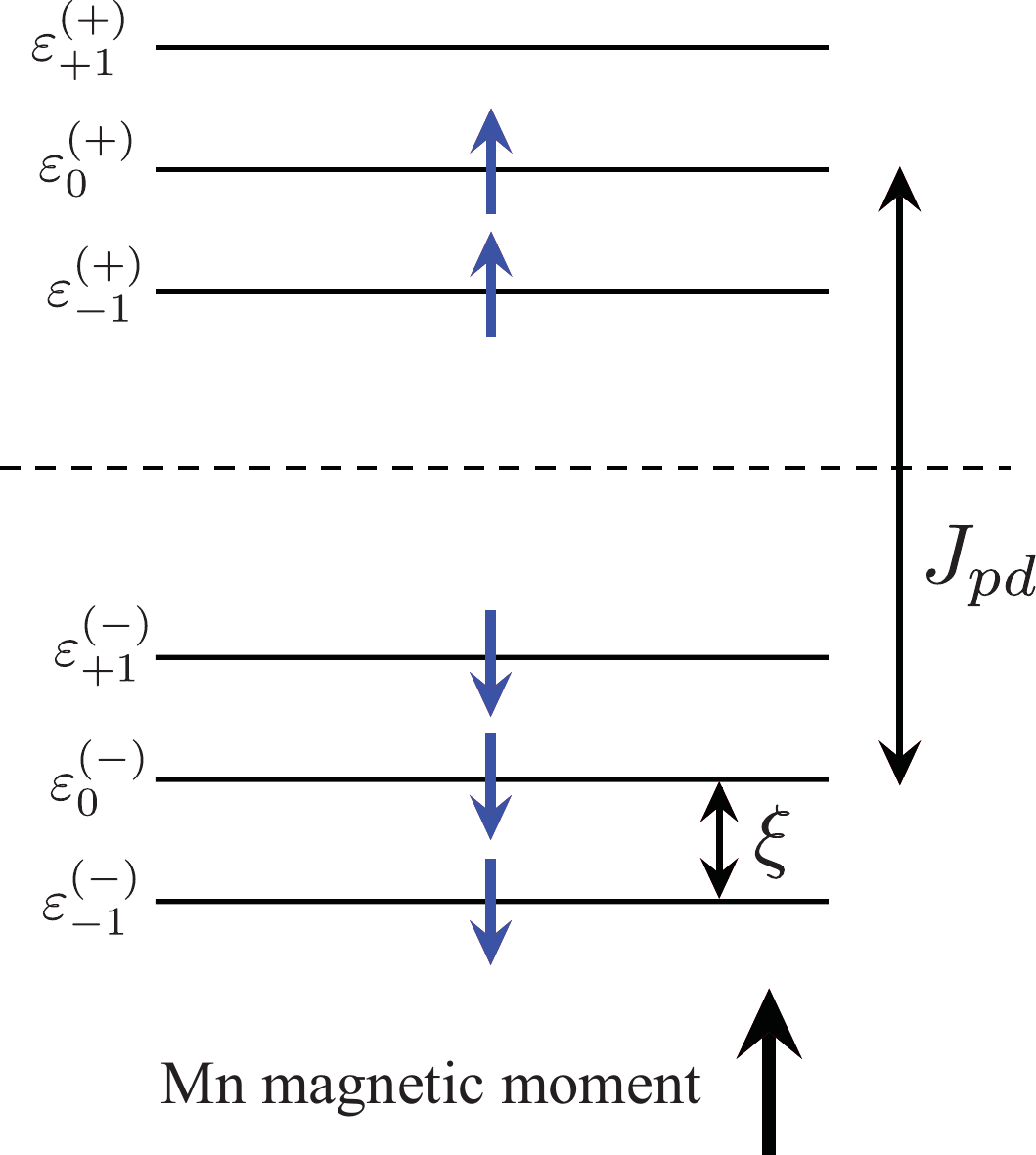}}\caption{The six
spin-polarized levels resulting from the $p$-$d$ hybridization around a Mn
impurity, shown in Fig.~\ref{cartoon2}. The three levels with orbital index
$\mu=-1,0,+1$ above the valence band edge, polarized in the direction of the
Mn moment, are higher in energy than the three corresponding levels polarized
in the opposite direction by an amount of the order of the effective exchange
constant $J_{pd}$. Spin-orbit interactions (and surface effects when present)
lift the orbital degeneracy of the like-spin levels and cause a small
admixture of opposite spin character. The topmost up-spin level is the
acceptor level introduced by the Mn.}%
\label{six_levels}%
\end{figure}

\begin{figure}[ptb]
\resizebox{7.3cm}{!}{\includegraphics{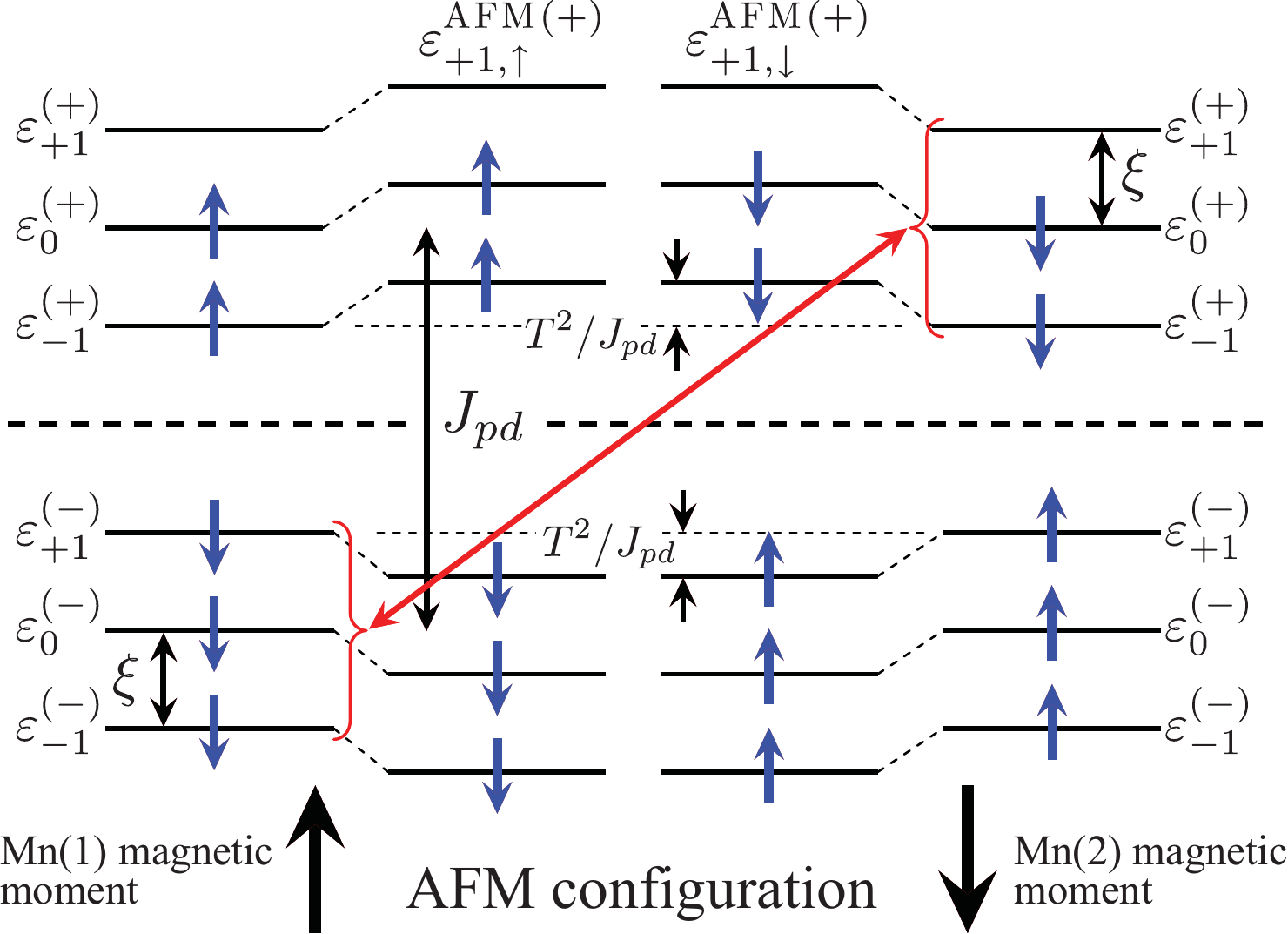}}\caption{Toy model for two
Mn atoms when the two local moments are aligned antiparallel -- the AFM
configuration. Each Mn site has six exchange-split itinerant levels that
hybridize with the like-spin levels at the other Mn site. Note that like-spin
levels have different energies at the two sites. The resulting hybridized
levels, drawn in the middle, have their energy lowered or raised by $\sim
T^{2}/J_{pd}$ where $T$ is the intersite hopping energy scale. This results in
a net gain in energy because of the two holes introduced by the two Mn ions.}%
\label{afm_pair}%
\end{figure}

\begin{figure}[ptb]
\resizebox{7.3cm}{!}{\includegraphics{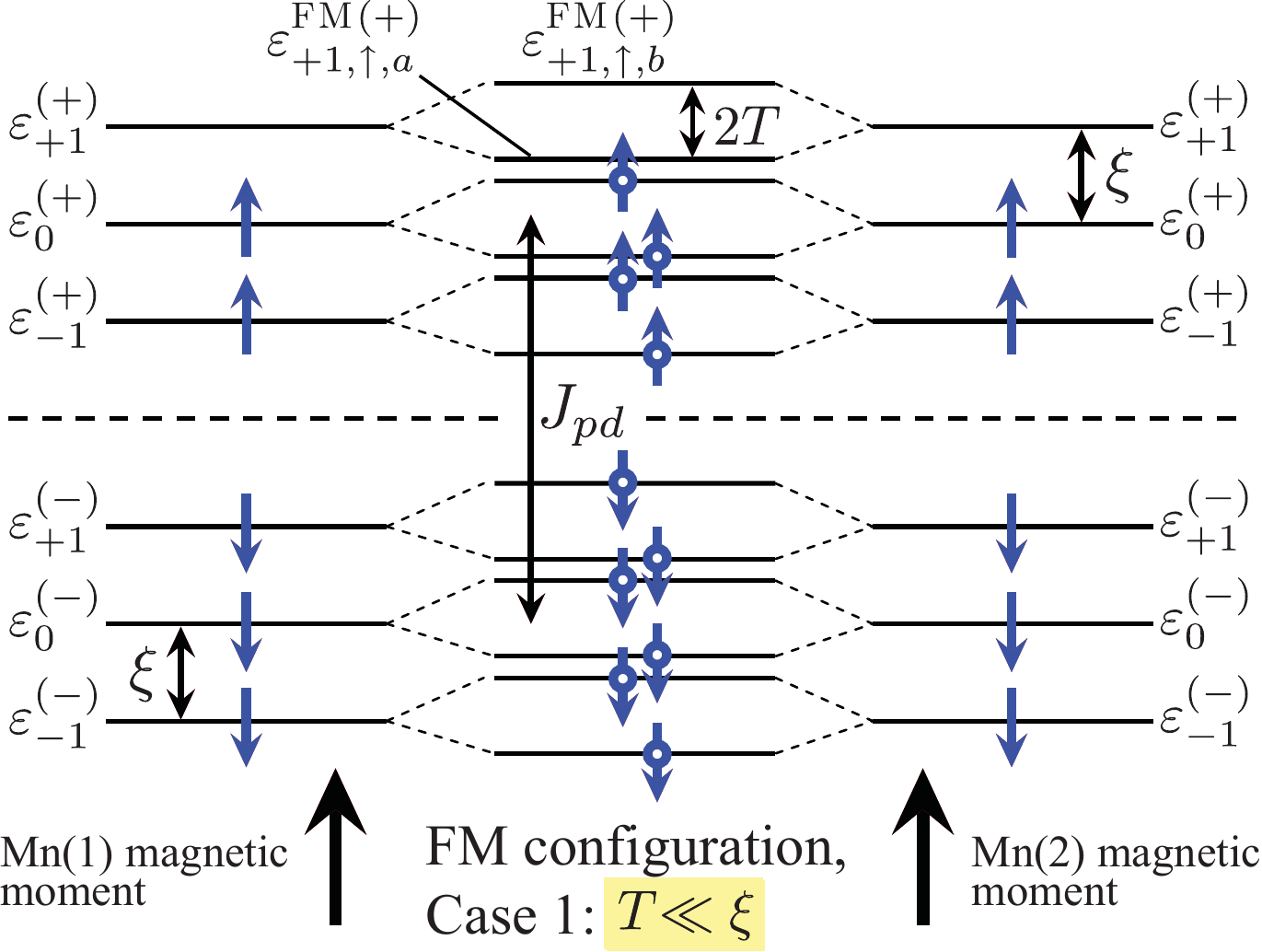}}\caption{Toy model for two
Mn atoms when the two local moments are aligned parallel -- the FM
configuration. Each Mn has six exchange-split itinerant levels that hybridize
with the like-spin levels at the other Mn via hopping. Now, like-spin levels
at the two sites are degenerate and give rise to bonding and antibonding
states split by the hopping parameter $2T$ which causes a \textquotedblleft
widening of the band\textquotedblright. The resulting hybridized levels are
drawn for the case when the hopping parameter $T$ is smaller than the
spin-orbit induced splitting $\xi$. Note that for the acceptor (hole) states
above the valence band edge, we use the convention of calling
\textquotedblleft bonding\textquotedblright\ the state with the higher energy
(subscript $b=$ bonding and $a=$ antibonding).}%
\label{fm_pair_1}%
\end{figure}

\begin{figure}[ptb]
\resizebox{7.3cm}{!}{\includegraphics{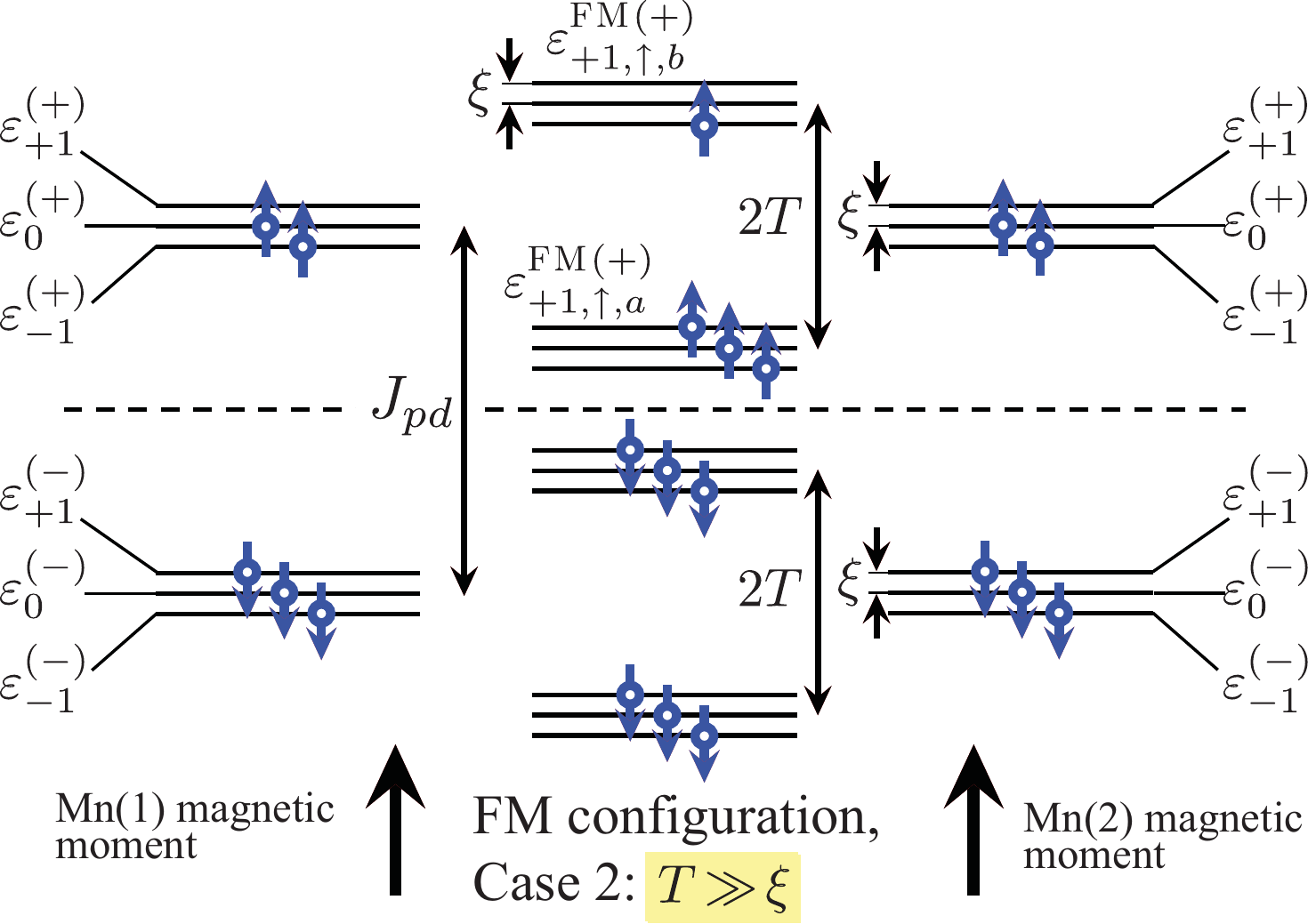}}\caption{The same FM
alignment of Mn spins as in Fig.~\ref{fm_pair_1}, but for the case when the
hopping parameter $T$ is much larger than the spin-orbit induced splitting
$\xi$. The resulting \textit{impurity band}\cite{endnote50}, of bandwidth
$\approx2T$, is partially filled when two holes are present, resulting in a
large energy gain with respect to the uncoupled $T=0$ system. }%
\label{fm_pair_2}%
\end{figure}

\subsection{Toy model for Mn pairs in GaAs}

\label{toy_model} In this section we discuss a system of two Mn atoms in GaAs
by means of a simple toy model\cite{schilf_prb01} that elucidates the basic
mechanism responsible for the effective coupling between their localized
spins. \footnote{A simplified version of this model was considered in
Ref.~[\onlinecite{schilf_prb01}]. Essentially the same model was used in
Ref.~[\onlinecite{yazdani_nat06}] to interpret the experimental results. Here
we consider a slightly more generalized version and we look in detail its
properties in different regimes of the parameters defining the model.} The
properties of this simple model will be very useful in interpreting the
results of our numerical calculations in Sec.~\ref{results}.

For each Mn impurity in GaAs we will focus on the six spin-polarized levels at
the top of the valence band, emerging from the $p$-$d$ hybridization shown in
Fig.~\ref{cartoon2}. These six levels are shown explicitly in
Fig.~\ref{six_levels} for one particular orientation of the Mn moment. The
three levels above the valence-band edge with orbital angular momentum
projection $\mu=0,\pm1$, are spin-polarized in the direction of the Mn
magnetic moment. Similarly, the corresponding three levels below the valence
band edge are polarized in the opposite direction. Spin-orbit coupling (and
surface effects when the Mn is not in the bulk) lifts the orbital degeneracy
of the like-spin states. We model this splitting by introducing the energies
\begin{equation}
\varepsilon_{\mu}^{(\pm)}=\pm J_{pd}/2+\mu\xi\;,\qquad\mu=0,\pm1\,
\end{equation}
where $\xi$ represents the SO-induced splitting. Here $(+)$ and $(-)$ stand
for the levels above and below the valence-band edge, respectively. Note that
$\varepsilon_{\mu}^{(+)}-\varepsilon_{\mu}^{(-)}=J_{pd}$, where $J_{pd}$ is
the effective exchange coupling used in Eq.~(\ref{hamiltonian}). In the
presence of spin-orbit interaction the spin is no longer a good quantum
number, but we will assume that these states still have a predominant spin
character, which is the same as when SO interaction is absent.

When two Mn impurities are present and close to each other, the system can
lower the total energy by allowing hopping between two single-particle states
with the \textit{same spin}, each centered around one of the two Mn. Two
different situations arise depending on whether the relative orientation of
the two Mn moments is parallel or antiparallel. We assume that each state
$|i,\mu,\sigma\rangle$, with $\sigma=\uparrow,\downarrow$ at Mn site $i$ will
be coupled to the corresponding same spin state $|j,\mu,\sigma\rangle$ at the
other Mn site $j$ by an effective hopping parameter $T_{\mu}$, which we take
to be spin-independent. The single-particle Hamiltonian representing these two
sets of itinerant spins coupled to the Mn local moment with hopping between
sites is given by
\begin{align}
H_{AFM}^{FM}=\sum_{\mu}  &  \Big(\varepsilon_{\mu}^{(+)}c_{1\mu\uparrow}%
^{\dag}c_{1\mu\uparrow}^{\phantom{\dag}}+\varepsilon_{\mu}^{(-)}%
c_{1\mu\downarrow}^{\dag}c_{1\mu\downarrow}^{\phantom{\dag}}\nonumber\\
&  +\varepsilon_{\mu}^{(\pm)}c_{2\mu\uparrow}^{\dag}c_{2\mu\uparrow
}^{\phantom{\dag}}+\varepsilon_{\mu}^{(\mp)}c_{2\mu\downarrow}^{\dag}%
c_{2\mu\downarrow}^{\phantom{\dag}}\Big)\nonumber\\
&  -\sum_{i\neq j,\sigma}T_{\mu}c_{i\mu\sigma}^{\dag}c_{j\mu\sigma
}{\phantom{\dag}}, \label{toyham}%
\end{align}
where the upper signs in superscript refer to the FM and the lower signs to
the AFM configuration.\footnote{In general, the effective hopping parameters
are also off-diagonal in orbital index. In the spirit of the qualitative
description of the toy model we disregard this complication.}
The Hamiltonian in Eq.~(\ref{toyham}) corresponds to the Anderson-Hasegawa
Hamiltonian\cite{anderson_pr55} with the angle $\theta$ between the spins
equal to 0 and $\pi$, for the parallel and antiparallel configurations respectively.

The Hamiltonian is immediately diagonalized by noting that the spin and
orbital characters are good quantum numbers. The nature of the resulting
hybridized levels will depend on the relative orientation of the two Mn
moments: in the FM configuration the two sets of unperturbed equal-spin states
(one at each Mn site) are degenerate and will form bonding and antibonding
combinations via hopping. By contrast, in the AFM configuration like-spins at
different Mn sites have different energies and the hybridization will be
reduced by a factor $\approx T_{\mu}/J_{pd}$.

In the AFM configuration the spectrum is doubly degenerate with energies
\begin{equation}
\varepsilon_{\mu\uparrow}^{\mathrm{AFM}(\pm)}=\varepsilon_{\mu\downarrow
}^{\mathrm{AFM}(\pm)}=\mu\,\xi\pm\sqrt{T_{\mu}^{2}+J_{pd}^{2}/4}.
\end{equation}
Fig.~\ref{afm_pair} shows a schematic view of the energy levels for the AFM
configuration with a constant $T_{\mu}=T$. In the presence of holes, the total
energy of the AFM state, $E_{\mathrm{tot}}^{\mathrm{AFM}}$, obtained by
summing the energies of the occupied states, is lower than for the
non-hybridized ($T_{\mu}=0$) state. The maximum gain occurs for the
\textquotedblleft half-filled\textquotedblright\ system consisting of six
holes. For the case of two Mn introducing two holes (shown in
Fig.~\ref{afm_pair}), the energy gain of the AFM configuration is
approximately $2T^{2}/J_{pd}$. This phenomenon is similar to the superexchange
mechanism -- arising within a one-band Hubbard model at half filling -- which
favors antiferromagnetic alignment of the itinerant spins on neighboring
sites. Note that there are no $d$ electrons present in our model and that the
superexchange between the local moments is brought about by the kinetic
exchange between the itinerant spins. When the two Mn atoms are close to each
other, oppositely aligned Mn spins allow the wave functions to spread out,
thus lowering their kinetic energy by hopping.

In the FM configuration the spectrum is nondegenerate
\begin{align}
&  \varepsilon_{\mu\uparrow b}^{\mathrm{FM}(+)}=\varepsilon_{\mu}^{(+)}%
+T_{\mu}\;,\quad\varepsilon_{\mu\uparrow a}^{\mathrm{FM}(+)}=\varepsilon_{\mu
}^{(+)}-T_{\mu}\;,\label{fm_energies_up}\\
&  \varepsilon_{\mu\downarrow a}^{\mathrm{FM}(-)}=\varepsilon_{\mu}%
^{(-)}+T_{\mu}\;,\quad\varepsilon_{\mu\downarrow b}^{\mathrm{FM}%
(-)}=\varepsilon_{\mu}^{(-)}-T_{\mu}\;,\label{fm_energies_down}%
\end{align}
where $\varepsilon_{\mu(\uparrow,\downarrow)b}^{\mathrm{FM}}$ and
$\varepsilon_{\mu(\uparrow,\downarrow)a}^{\mathrm{FM}}$ are bonding and
antibonding states respectively. The acceptor (hole) states are more bound
when they are further away from the valence band top, which is the start of
the continuum for hole states. Out of the two acceptor states arising from the
hybridization of the degenerate like-spin states of energy $\varepsilon_{\mu
}^{(+)}$, we therefore assign\footnote{Within the toy model, this choice of
labeling the states of higher energy as \textquotedblleft
bonding\textquotedblright\ is just a convention, since we do not really have a
continuum from the valence band. However we will see that this convention is
consistent with the results of the numerical calculations for the real system,
where a valence band quasi-continuum does exist.} the label \textquotedblleft
bonding\textquotedblright\ to the one that occurs at higher energy, while the
one with lower energy is denoted as \textquotedblleft
antibonding\textquotedblright\ [see Eq.~\ref{fm_energies_up}].

The energy spectrum for the FM configuration is shown in Fig.~\ref{fm_pair_1}
when the hopping parameter is smaller than the SO splitting, $T_{\mu}\equiv
T\ll\xi$. The six$\frac{{}}{{}}$ acceptor levels above the valence band edge
form an \textquotedblleft impurity band\textquotedblright\cite{endnote50}%
\ with an associated bandwidth that increases with the splitting $T$. Note
that for $T\ll\xi$, when two holes are present the total energy
$E_{\mathrm{tot}}^{\mathrm{FM}}$ is the same as for the non-hybridized case,
$T=0$. An energy gain equal to $-T$ occurs only when an odd number of holes
are present. The opposite limit of a large hopping parameter, $T_{\mu}\equiv
T\gg\xi$, is shown in Fig.~\ref{fm_pair_2}. Since only two thirds of the
\textquotedblleft impurity band\textquotedblright\ is filled there is a net
energy gain $\xi-2T$ due to the level splitting, which can stabilize the FM
configuration against the AFM one. This mechanism that couples the itinerant
spins ferromagnetically, corresponds to double exchange between the two sites.
For a partially occupied impurity band the FM alignment tends to be
energetically more favorable than the AFM when the widening of the band
induced by hopping is large. The difference between the total energies of the
two configurations defines an effective exchange constant $J\equiv
(E_{\mathrm{tot}}^{\mathrm{AFM}}-E_{\mathrm{tot}}^{\mathrm{FM}})/2$
representing a Heisenberg-like magnetic interaction, $\mathcal{H}\propto
-J\vec{S}_{\text{Mn}(1)}\cdot\vec{S}_{\text{Mn}(1)}$, between the local
moments at the two Mn sites.

\begin{figure}[ptb]
\resizebox{7.5cm}{!}{\includegraphics{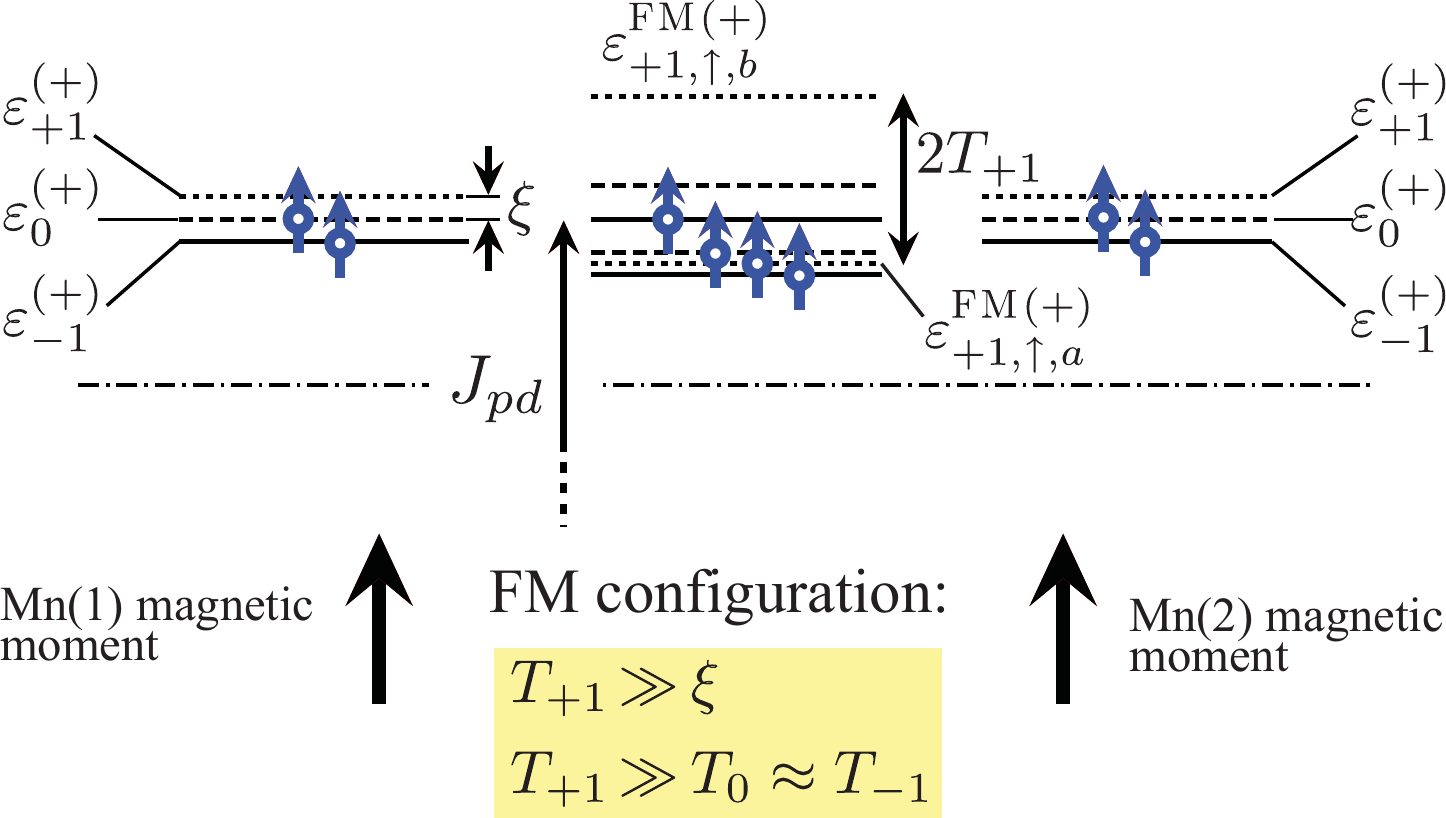}}\caption{Toy model for two
Mn sites in the FM configuration when one hopping matrix element ($T_{\mu=+1}%
$) is much larger than the other two and larger than the spin-orbit induced
splitting $\xi$. Only the impurity levels above the valence-band edge are
drawn. The half-occupied impurity band favors a FM alignment over the AFM
configuration and the non-hybridized ($T_{\mu}=0$) system. The acceptor
splitting is large and equal to $T_{\mu=+1}-T_{\mu=0}+\xi$. As we will see
later (following the discussion of Fig.~\ref{bulklevels}), this is the case
that best describes the numerical results of the microscopic Hamiltonian in
Eq.~(\ref{hamiltonian}). }%
\label{fm_pair_3}%
\end{figure}

The actual hopping paths are of course more complicated than the ones shown in
Figs.~\ref{afm_pair}-\ref{fm_pair_2}. In particular, for a hopping parameter
$T_{\mu}$ depending strongly on the orbital character $\mu$, we expect the
resulting FM configuration to be intermediate between the limiting cases of
Fig.~\ref{fm_pair_1} and Fig.~\ref{fm_pair_2}. In any case, the toy model
predicts that the FM configuration will always be characterized by a splitting
$\Delta_{\mathrm{acc}}$ of the two acceptor states, related either to
covalency between individual acceptor levels or to spin-orbit coupling. The
splitting is noticeably absent in the AFM configuration, where the two
acceptor states are quasi-degenerate, and is therefore a landmark of the FM
state. As mentioned in the introduction, the STM experiments do measure a
significant splitting between the two acceptor states, which is a strong
indication that the Mn pair is coupled ferromagnetically. It is therefore
important to investigate if some kind of relationship exists between
$\Delta_{\mathrm{acc}}$ and $J$. This can be the case if, for example, the
level structure of the FM configuration is of the type sketched in
Fig.~\ref{fm_pair_3}. Here a dominating hopping term $T_{\mu=+1}$ gives rise
to both a partially filled impurity band\cite{endnote50} stabilizing the FM
state, and a large acceptor splitting $\Delta_{\mathrm{acc}}=T_{\mu=+1}%
-T_{\mu=0}+\xi$, which will also be approximately related to the exchange
energy gained by the FM configuration. As we will see later (following the
discussion of Fig.~\ref{bulklevels}), this is the case that best describes the
numerical results of the microscopic Hamiltonian in Eq.~(\ref{hamiltonian}).

In the next Section we examine the properties of the six impurity levels when
the Mn spins are parallel or antiparallel, and see how these states relate to
the effective exchange constant $J$ and the magnetic anisotropy energy. Note
that the trace of the $p$-$d$ exchange operator, summed over all valence band
states is zero. If the impurity levels involve negligible conduction band
character, it follows that the sum of the energies of all valence band states
is independent of Mn spin orientations. \cite{scm_MnGaAs_paper1_prb09} Because
of this property, exchange and interaction energies are expected to be
accurately expressed in terms of the energies of the two empty valence band states.

In the evaluation of $J$, we will also approximate the total energy for the FM
and AFM configurations by summing up the energies of the two unoccupied
acceptor levels, or alternatively of the four topmost occupied levels. These
results are compared the $J$ obtained by the difference between AFM and FM
total energies. We will also consider other measures of the acceptor level
structure broadening in Mn dimers, which might be more closely related to
exchange interaction than the splitting of the top two levels. These measures
include the splitting between the mean of the four occupied and the mean of
the two unoccupied levels, and the effective "bandwidth" of the six levels as
obtained by calculating the standard deviation from their mean value.

\section{Results}

\label{results}

\begin{figure}[ptb]
\resizebox{6.5cm}{!}{\includegraphics{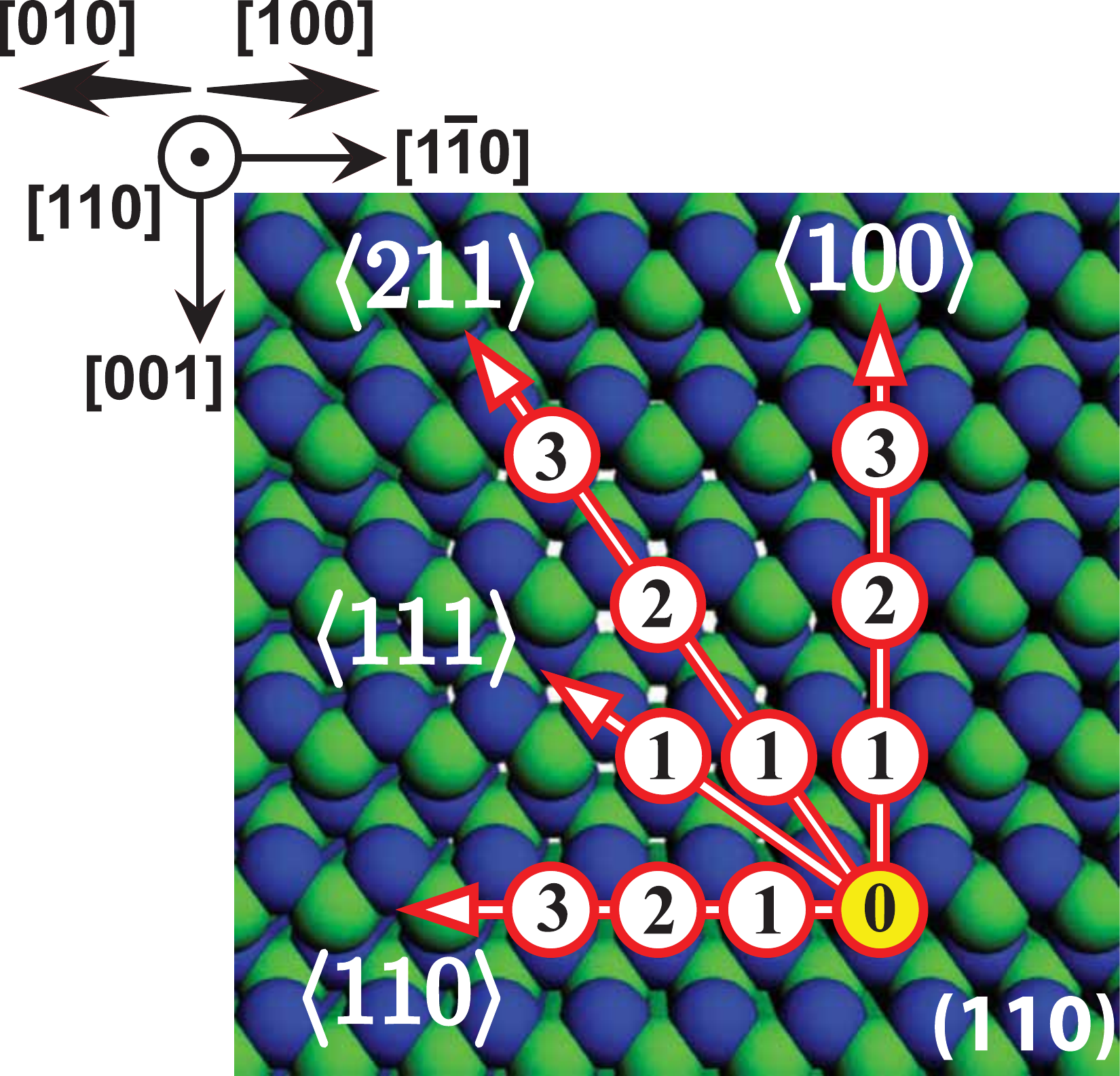}}\caption{(Color online)
Directions and separations of a Mn pair on the (110) surface. The two Mn
atoms, replacing Ga atoms (bright green spheres), are marked 0 (first Mn atom)
and 1,2,3 (second Mn atom) in correspondence with Fig.~\ref{fmae} below. Note
that all equivalent symmetry directions $\langle...\rangle$ have been chosen
in direct coordinates $[...]$, such that all the Mn pairs appear in the
$\left(  110\right)  $ plane. }%
\label{pair_directions}%
\end{figure}

\subsection{Mn-Mn interactions in bulk GaAs}

\label{bulkresults}

In this section we study numerically the properties of substitutional Mn pairs
in \textit{bulk} GaAs. As shown in Fig.~\ref{pair_directions}, we consider Mn
pairs oriented along different crystalline directions at various separations.
The two Mn are embedded in a 3200 atom GaAs supercell with periodic boundary
conditions in all directions, corresponding to a Mn fraction of 0.06\%. We
consider collinear magnetic configurations in which the Mn moments are either
parallel or antiparallel. \begin{figure}[ptb]
\resizebox{7.5cm}{!}{\includegraphics{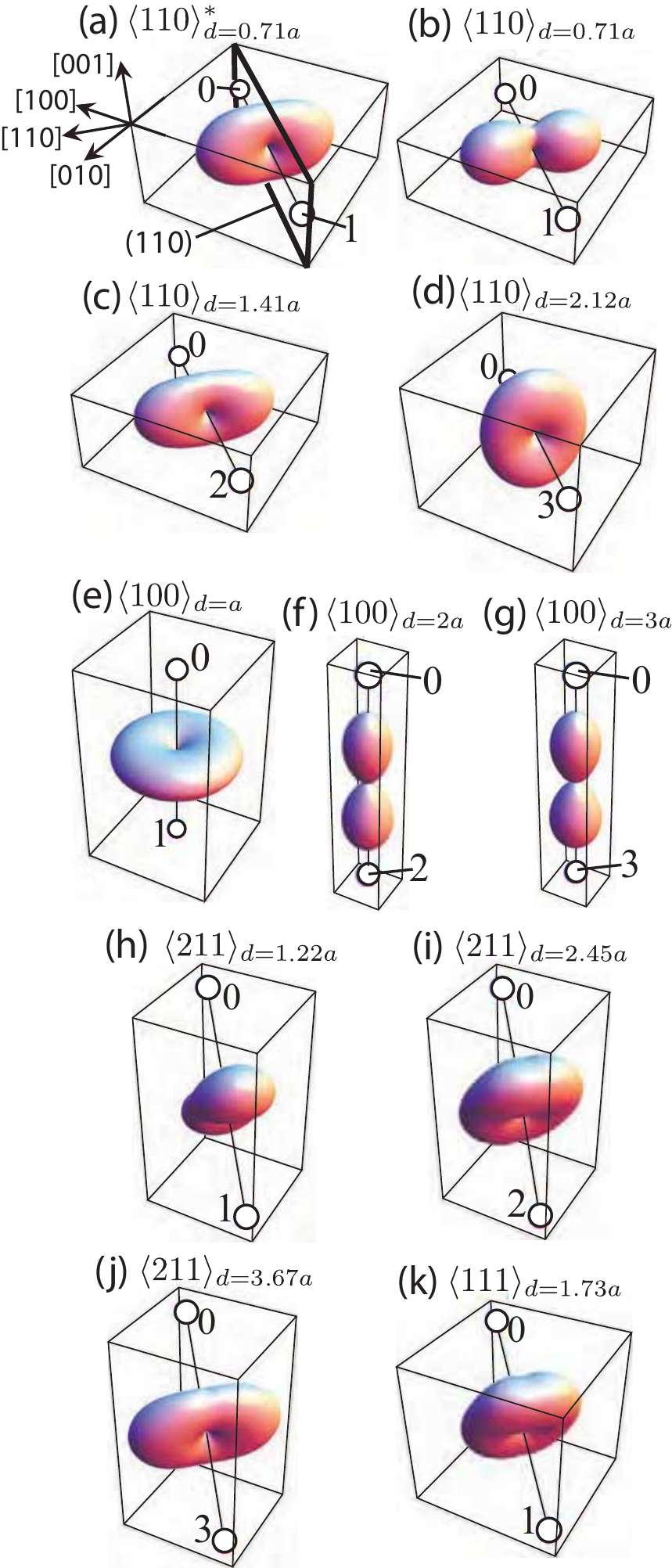}}\caption{(Color online)
Magnetic anisotropy in collinear variation of parallel Mn spins on the unit
sphere. Generally, an easy axis is found to be parallel to the connecting Mn
line, with the exception of the $\langle100\rangle_{d=2a,3a}$. The
$\langle110\rangle^{\ast}$ shows the effect of non-additive Coulomb correction
on the common As neighbor, yielding larger anisotropy barriers. In panel (a)
the crystal axes have been marked and the same labels apply to all other
panels. Also indicated is the (110) plane that contains all Mn pairs. The Mn
are numbered in accordance with Fig.~\ref{pair_directions}. The AFM
configurations all have a higher total energy and the variations are inverted
with respect to the FM ones. }%
\label{fmae}%
\end{figure}

\subsubsection{Magnetic anisotropy energy}

\label{bulk_anysotropy} We begin by looking at the magnetic anisotropy of the
system, which is defined as the dependence of the total ground-state energy
$E_{\mathrm{tot}}(\hat{\Omega})$ on the direction of the Mn-pair magnetic
moment, $\hat{\Omega}$. For the AFM configuration, $\hat{\Omega}$ is the
direction of the staggered moment. Graphical representations of the magnetic
anisotropy landscapes $E_{\mathrm{tot}}(\hat{\Omega})$ on the unit sphere of
the all possible directions for $\hat{\Omega}$, are presented in
Fig.~\ref{fmae} for the FM configurations. Each panel (a)-(k) refers to a Mn
pair with a particular separation and orientation in the crystal, according to
the notation defined in Fig.~\ref{pair_directions}.\begin{figure}[ptb]
\resizebox{6.5cm}{!}{\includegraphics{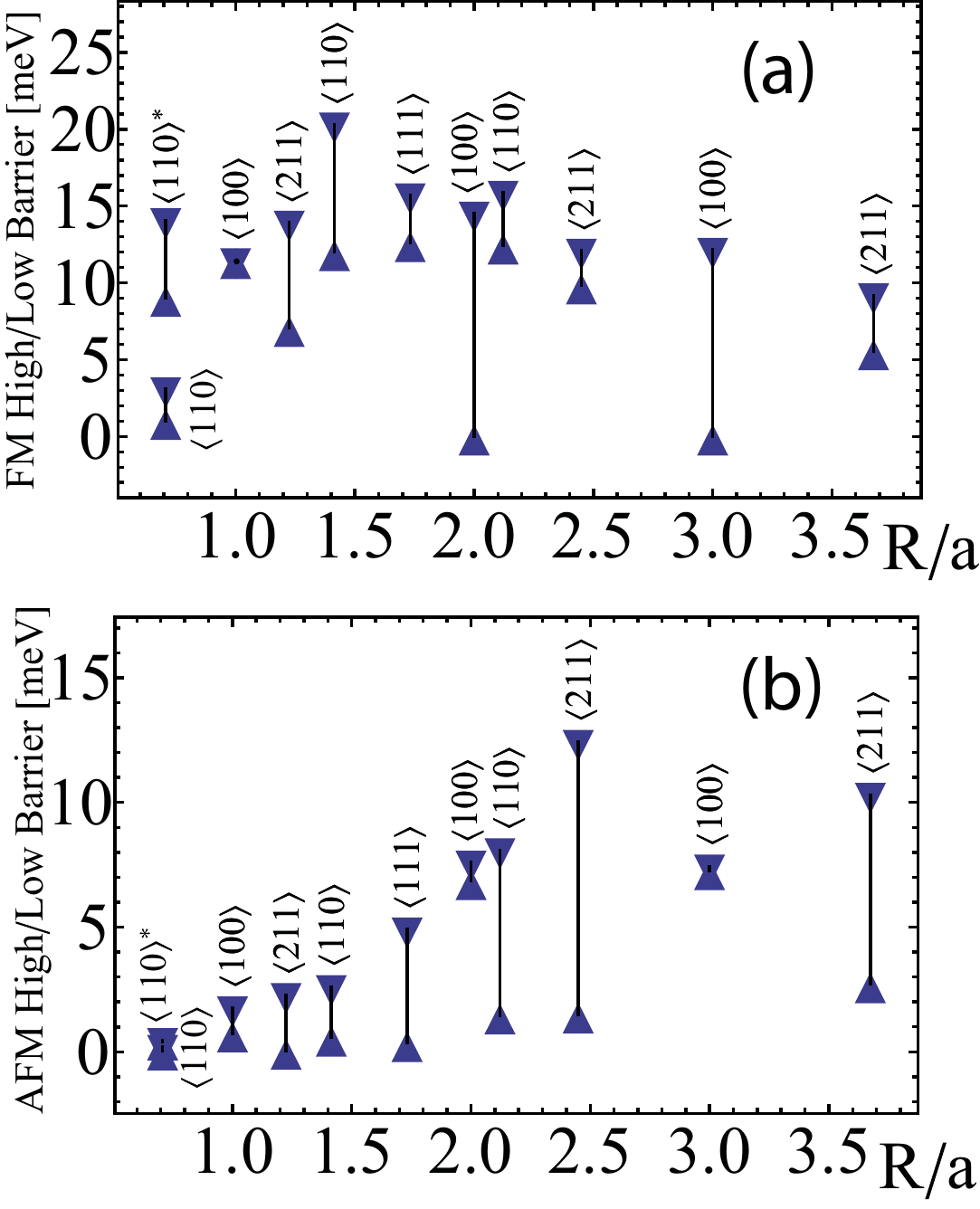}}\caption{High and Low
magnetic anisotropy barriers. In these graphs, zero corresponds to the minimum
energy on the unit sphere. The anisotropy increases when the Coulomb
correction is taken to be non-additive for $\langle110\rangle^{\ast}$ in the
FM configuration (a). The $\langle110\rangle_{d=1.41}$ shows enhanced
anisotropy barriers, with strong interactions along this symmetry direction.
$\langle100\rangle$ has a single barrier for the closest distance, but
exhibits quasi-easy-planes (low barrier is zero) perpendicular to the
connecting Mn line for longer distances. In the AFM configurations (b), the
anisotropies gradually increase with distance as cancellations between the two
acceptor levels decrease.}%
\label{highlow}%
\end{figure}In these figures, what we actually plot is
\begin{equation}
E_{\mathrm{anis}}(\hat{\Omega})\equiv E_{\mathrm{tot}}(\hat{\Omega
})-E_{\mathrm{tot}}^{\mathrm{min}}, \label{eanis}%
\end{equation}
as a function of the magnetic moment direction $\hat{\Omega}$. Each point of
the \textquotedblleft anisotropy surface\textquotedblright\ is obtained by
measuring the "distance" $\big(E_{\mathrm{tot}}(\hat{\Omega})-E_{\mathrm{tot}%
}^{\mathrm{min}}\big)$ from the center of the reference parallelepiped defined
by the cubic axes $[100]$, $[010]$, and $[001]$. The center of the
parallelepiped is also taken to be the center of the unit sphere of
directions. Here $E_{\mathrm{tot}}^{\mathrm{min}}$ is the minimum value of
$E_{\mathrm{tot}}(\hat{\Omega})$ upon varying $\hat{\Omega}$ that occurs along
one of the easy directions. The actual values of the high and the low energy
barriers of these systems are shown in Fig.~\ref{highlow}. In the FM
configurations, the bistable easy directions are generally found to be
parallel to Mn pair axis. The two exceptions are the quasieasy planes
perpendicular to the connecting line, formed when the Mn are spaced at two and
three lattice constants along $\langle100\rangle$ [see Fig.~\ref{fmae} (f) and
(g)]. For these two configurations, the magnetic hard direction has 12-15 meV
higher energy. The qualitatively different landscapes signal that the
interactions along the $\langle100\rangle$ direction differ from the other
crystalline directions. For the shortest separation along $\langle100\rangle$,
$d=a=5.65~$A, a bistable easy axis parallel to the Mn-connecting line with a
blocking barrier of around 11 meV is found. This means that a level crossing
at the Fermi level occurs when the distance is increased from one to two
lattice constants. Associated with this crossing is a change in the orbital
character of the acceptors that results in a qualitatively different anisotropy.

Focusing now on the $\langle110\rangle$ [with the Mn pair along $[1\bar{1}0]$,
see Fig.~\ref{fmae} (b)-(d)], for which closely spaced As and Ga provide more
direct hopping paths between the Mn, we see that there is a low barrier along
the [001] axis and a high barrier along [110]. At the closest spacing,
$d=0.7a$, the anisotropy energy is very small with barriers of 1-3 meV [see
Fig.~\ref{fmae} (b)]. This configuration is special, because the off-site
Coulomb correction on the common nearest neighbor As between the Mn is
additive, giving it a large on-site energy that reduces anisotropy. The
$\langle110\rangle^{\ast}$ pair in Fig.~\ref{fmae} (a) shows the effect of a
non-additive\footnote{Non-additive Coulomb correction means that the As
neighboring the two Mn, receives $V_{\mathrm{off}}$ rather than
$2V_{\mathrm{off}}$ in the $V_{\mathrm{corr}}$ term in Eq.$~$%
(\ref{hamiltonian}).} Coulomb correction on the common As neighbor, where the
low and high barriers have increased to 9 and 14 meV. As we let the Mn move
apart one step further along the $\langle110\rangle$ with distance $d=1.4a,$
the high barrier attains the maximum value of all considered configurations of
21 meV, and a low barrier of 12 meV. At the longest considered separation
along $\langle110\rangle$, the low and high barriers are now 12 and 16 meV,
respectively. The $\langle211\rangle$ Mn pairs follow a similar evolution of
the high and low barrier, relative the connecting Mn line, now with a low
barrier along the [110] direction that changes to the high barrier with
increasing distance. Distances are longer than for the $\langle110\rangle$
pairs and therefore hybridization is weaker and the anisotropy energies drops
by a few meV. At the larger separations along $\langle211\rangle$ the bistable
minima are tilted away from the connecting line, an indication that the single
Mn anisotropy is becoming comparable to the Mn-Mn interaction energy. Even
with a supercell of 3200 atoms we cannot exclude possible finite-size-effect
contributions to the anisotropies. Nevertheless, we can get a good estimate of
the barriers separating the generally bistable minima. The magnetic anisotropy
in (Ga,Mn)As nanostructures, such as (Ga, Mn)As epilayers, is presently a
topic of great interest.

Understanding the microscopic mechanisms of magnetic anisotropy in DMS is
crucial in order be able to manipulate the magnetization vector by magnetic
and electric fields.\cite{chiba_nat08} For a review of recent theoretical and
experimental works on magnetic anisotropy in (Ga, Mn)As see Ref.
[\onlinecite{zemen_prb09}], where ferromagnetic samples with a relatively high
Mn content of a few percent are examined. Comparison with our results, which
instead pertain to the magnetic anisotropy of isolated Mn dimers, is not
straightforward. Our results could on the other hand be directly compared with
those of STM experiments using a spin-polarized magnetic
tip\cite{wiesendanger_rmp09} and an external magnetic field.

As we show below, all the AFM pairs have a higher total energy, which is
expected on the basis of the heuristic toy model considerations. The AFM
anisotropy landscapes are qualitatively inverted with respect to the FM ones,
i.e., the bistable minima along the Mn dimer axis in the FM variation are
replaced by quasieasy planes perpendicular to the dimer axis in the AFM
variation. This inversion of hard and easy directions between FM and AFM can
be understood by considering the FM configuration in the easy direction, where
the spin-orbit splitting between the highest occupied and the lowest
unoccupied levels is large, resulting in a lower total energy. In the
corresponding AFM configuration, the two levels are of opposing spin
character, such that the splitting is smaller and the total energy instead
higher. In Fig.~\ref{highlow} (b), we see that the closer the Mn are, the
smaller the anisotropy energies tend to be. This figure gives us an idea of
the range of strongly interacting Mn. Weaker interactions at larger distances
in the AFM configuration tends to increase the anisotropy, as the effectively
spin-polarized region in the lattice around each Mn increases.

\subsubsection{Character of acceptors for Mn pairs in bulk GaAs}

\label{bulk_acceptor}The anisotropy of the embedded Mn dimers is accurately
reflected in the anisotropy of the two acceptor levels and their associated
wave functions. In this section we therefore perform a detailed analysis of
various properties of the acceptor levels, such as their splitting and LDOS -
quantities that are directly accessible by STM spectroscopy.

From now on we use the convention to denote with $\varepsilon_{-3}%
\;,\varepsilon_{-2}\;,\varepsilon_{-1}\;,\varepsilon_{0}$ the energies of four
highest occupied states, while $\varepsilon_{1}$ and $\varepsilon_{2}$ denote
the energies of the two acceptor levels. When mixing with the conduction band
is negligible, the anisotropy energy landscape can be extracted from the the
two acceptor levels using%
\begin{equation}
E_{\mathrm{anis}}(\hat{\Omega})=-\sum_{i=1}^{2}[\varepsilon_{i}(\hat{\Omega
})-\varepsilon_{i}^{\mathrm{min}}], \label{accanis}%
\end{equation}
where $E_{\mathrm{anis}}(\hat{\Omega})$ is defined in Eq.~(\ref{eanis}) and
the upper limit in the sum can be generalized to a greater number of acceptor
levels. Equation (\ref{accanis}) follows from the fact that the trace of the
$p$-$d$ exchange operator in Eq.~(\ref{hamiltonian}) summed over all valence
band states including the acceptors is zero. In the FM variation of the Mn
spins, the high-energy acceptor level, $\varepsilon_{2}(\hat{\Omega})$, varies
very little and in a qualitatively opposite manner to the low-energy acceptor,
$\varepsilon_{1}(\hat{\Omega})$. Its effect is therefore to reduce the much
higher anisotropy coming from the lower acceptor level. In the AFM
configuration small acceptor level variations partially cancel to a low total
anisotropy below 1.5 lattice constants, and add up to a larger total
anisotropy above this value.

SO interaction mixes spin components, such that the eigenstates are no longer
of definite spin character. As a result, the acceptor states above the valence
band edge, which without SO coupling have the same spin character as the
corresponding localized Mn moment ("spin-up"), acquire a small component in
the opposite direction (\textquotedblleft spin-down\textquotedblright). In the
FM cases, this results in a small spin down character of the acceptor levels,
that increases with Mn distance primarily for the low acceptor, as it moves
closer to the valence band. In the AFM variations, the two acceptor levels are
now very close in energy and their spin character can vary by a large amount
on the unit sphere due to spin-orbit interaction. \begin{figure}[ptb]
\resizebox{6.5cm}{!}{\includegraphics{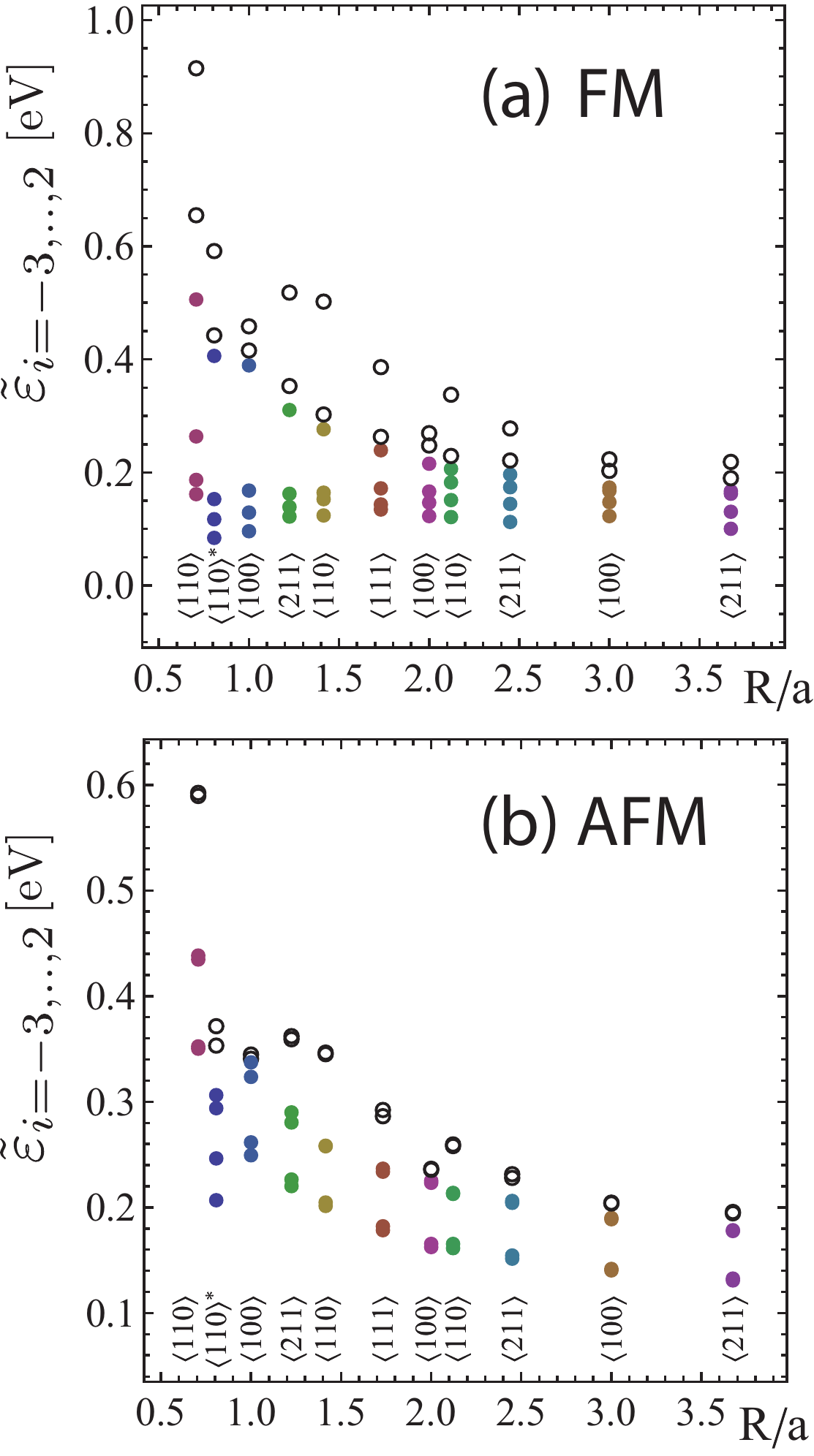}}\caption{The four highest
occupied (filled circles) and the two lowest unoccupied (empty circles)
eigenstates as obtained by taking the spherical average. In the FM
configuration (a) the splitting between the three upper and lower impurity
levels decreases with distance, but the split between the two acceptor levels
follows a more complex pattern, depending on pair orientation. In the AFM
configuration (b) levels bunch up but are still split from each other by
weaker hybridization. The $\langle110\rangle^{\ast}$ has been shifted slightly
to the right in order not to obscure the graph.}%
\label{bulklevels}%
\end{figure}

In Fig.~\ref{bulklevels} we plot the energies of the four highest occupied and
the two unoccupied acceptor levels for all studied configurations. We consider
spherical averages of these six energies over all possible moment directions,
denoted by $\tilde{\varepsilon}_{i}\;,i=-3\;,\dots\;,2$.

In the FM configuration [Fig.~\ref{bulklevels} (a)] we find that there is a
clear splitting between the three lowest and highest levels, in agreement with
the results of the toy model shown in Fig.~\ref{fm_pair_2}. The fact that
three highest levels are also spread over a sizable energy range indicates
that, at least for the shortest Mn separation, we are in the regime described
by Fig.~\ref{fm_pair_3} of the toy model. Note in particular the large
splitting of the two acceptor levels, as expected for aligned Mn spins. The
level splitting and the overall width of the impurity band\cite{endnote50}
decreases with increasing Mn separation. Note also the strong dependence of
the acceptor splitting on the orientation of the Mn pair. In particular, the
FM acceptor splittings tend to be largest along the $\langle110\rangle$
direction. It is also clear that the $\langle100\rangle$ Mn pair behaves
differently. Even if the splitting is large between the three lower and three
upper levels, the splitting of the two acceptor levels is consistently smaller
than for the other orientations.

In the AFM configuration the structure of the six energy levels is quite
different and its salient features are nicely captured by the toy model result
of Fig.~\ref{afm_pair}. The levels are always essentially doubly degenerate;
the splitting between doublets can be large for the shortest Mn separation and
but it decreases quickly with distance and becomes of the order of the
expected SO-induced splitting. In contrast to the FM case, there is no visible
splitting between the acceptor levels. The only exception is the special case
of the $\langle110\rangle$ for the shortest Mn separation, where the six
impurity levels abruptly drop down towards the valence continuum as the
off-site Coulomb correction on the common As neighbor is decreased. Finally
note that the $\langle100\rangle$ pair sticks out with a more dense set of
levels also in the AFM configuration

\begin{figure}[ptb]
\resizebox{6.0cm}{!}{\includegraphics{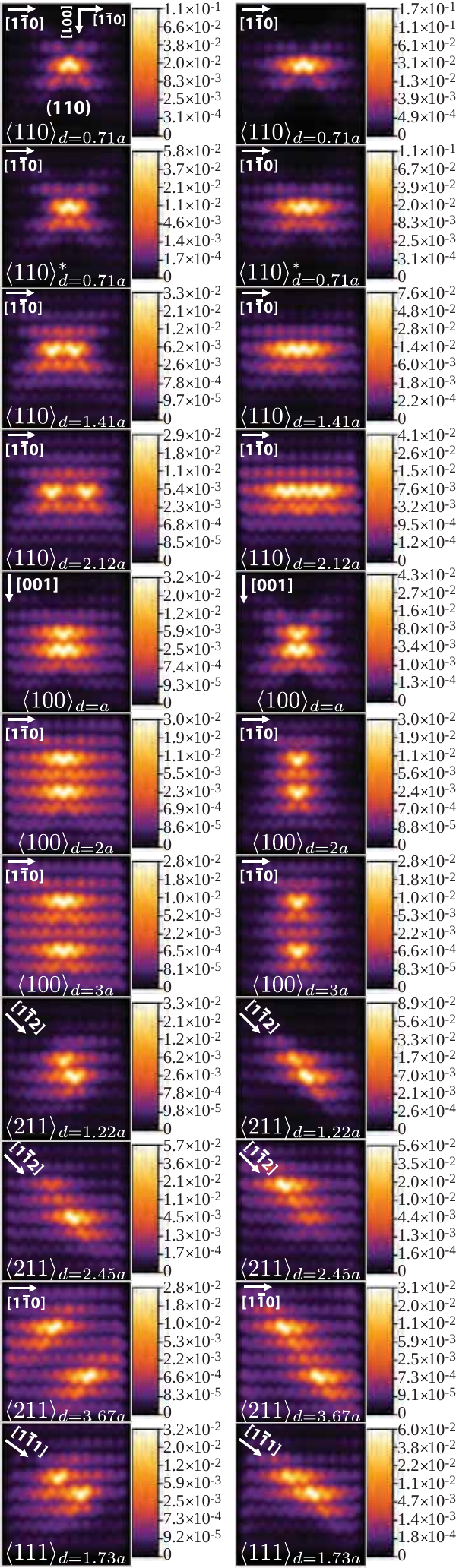}}\caption{(Color online)
LDOS of the lower-energy (left column) and the higher-energy (right column)
acceptor level in the easy direction (indicated by white arrow in top left
corners) for the FM configurations. Each row corresponds to a given crystal
orientation and separation for of the Mn pair. }%
\label{bulkldosfm}%
\end{figure}\begin{figure}[ptb]
\resizebox{6.0cm}{!}{\includegraphics{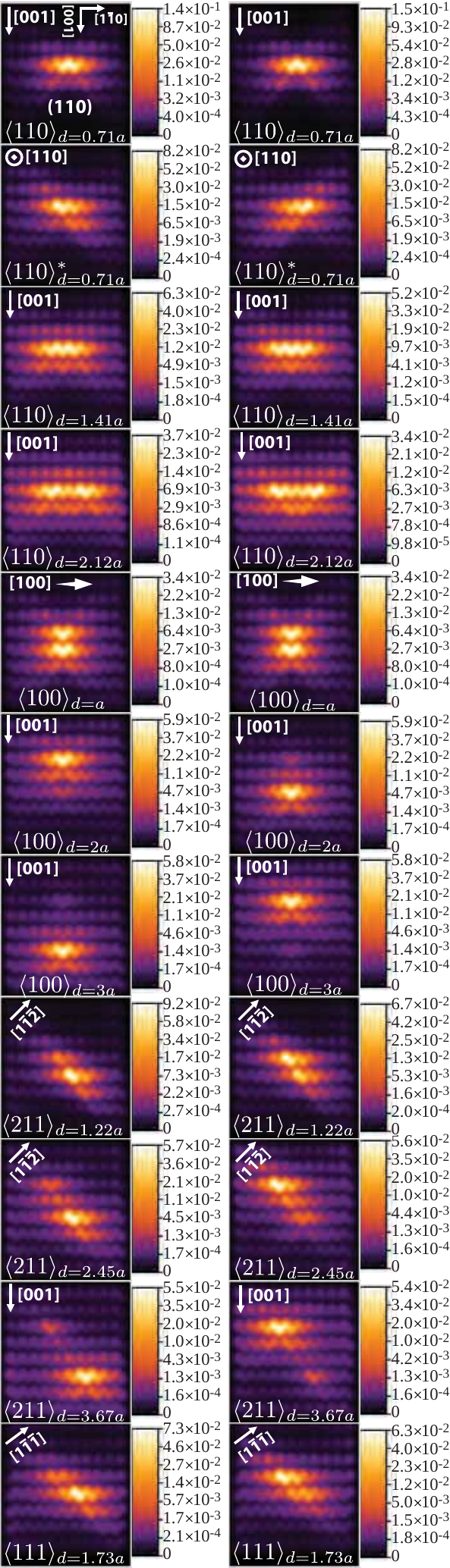}}\caption{(Color online)
LDOS of the lower-energy (left column) and the higher-energy (right column)
acceptor level in the easy direction for the AFM configurations.}%
\label{bulkldosafm}%
\end{figure}

We proceed to discuss the Local Density Of States (LDOS) of the two acceptor
levels - a property that can be probed by STM
spectroscopy.\cite{yazdani_nat06, yazdani_jap07,koenraad_prb08,yazdani_prb09}
Plots of the LDOS of the two acceptor levels in the (110) plane containing the
two Mn, are shown in Fig.~\ref{bulkldosfm} for the FM and in
Fig.~\ref{bulkldosafm} for the AFM configurations. The LDOS plots are
generated\footnote{Gaussians with a magnitude proportional to the calculated
lattice-model LDOS are placed at each atomic position in the (110) plane. The
Gaussians are given a full width at half maximum equal to half the nearest
neighbor distance in order to emulate the finite spatial resolution in the STM
experiments, and a logarithmic color scale is applied to facilitate comparison
with the constant-current-mode STM images.} from the tight-binding model
calculations as in
Refs.~[\onlinecite{scm_MnGaAs_paper1_prb09,tangflatte_prl04}]. In these plots
the Mn spins are pointing in the magnetic easy directions, indicated by the
arrow in the top left corner of each panel. The left column in each figure
shows the LDOS of the lower energy acceptor state, which is closer to the
valence band maximum; the right column refers to the acceptor with higher energy.

We first consider the $\langle110\rangle$ pairs in the FM configuration (see
Fig.~\ref{bulkldosfm}). For the special case where the Mn atoms are separated
by a common nearest neighbor As, the results change significantly depending on
whether the off-site Coulomb correction is additive ($\langle110\rangle
_{d=0.71a}$) or non-additive ($\langle110\rangle_{d=0.71a}^{\ast}$). We see
that the effect of a non-additive Coulomb correction on the common As is to
delocalize the acceptor wave functions. A more delocalized wave function
generates larger anisotropy energies, as shown in Fig.~\ref{highlow}. At the
shortest Mn separation, both acceptor wave functions have \textit{bonding
character}, with the maximum spectral weight on the common As neighbor located
between them. This situation is captured by the toy model, where the large
hybridization occurring at small Mn separation gives rise to the level
structure shown in Figs.~\ref{fm_pair_2} and \ref{fm_pair_3}, with both
acceptors being of the bonding type. As the Mn ions move apart along
$\langle110\rangle$, the lower acceptor state develops more
\textit{antibonding} character with a significant decrease of the spectral
weight between the two Mn sites, whereas the upper acceptor remains in a
bonding-like state. Within our toy model, this implies that the two acceptor
states correspond to bonding and antibonding states arising from the
hybridization of degenerate levels in energy, spin and orbital character, as
shown in Fig.~\ref{fm_pair_1}.

In agreement with our results, the STM experiments for the $\langle110\rangle$
pair find clear evidence that the two Mn-induced acceptor states have bonding
and antibonding character, with the bonding state occurring at higher
energies. \cite{yazdani_nat06, yazdani_jap07} An antibonding character for the
lower energy state is observed experimentally also for the Mn pair with the
shortest separation.\cite{yazdani_jap07} While this does not seem to be case
in our bulk $\langle110\rangle_{d=0.71a}$ calculations (both states being
essentially \textquotedblleft bonding\textquotedblright), we do find that the
value of the maximum LDOS on the As in between the Mn's is around twice as
large for the upper acceptor wave function.

For the $\langle100\rangle$ pairs the upper acceptor exhibits a more localized
signature, whereas the lower acceptor is more extended because it is closer to
the valence continuum. Bonding and antibonding formations cannot be clearly
seen, indicating that the hybridization between the two holes is much weaker
along the $\langle100\rangle$. This conclusion is also supported by the very
small acceptor-energy splitting, as shown in Fig.~\ref{surflevels}. In the
remaining directions $\langle211\rangle$ and $\langle111\rangle$ (rows 8-11),
bonding and antibonding characters are visible, with more and less spectral
weight in the region between the Mn, respectively.

In the AFM LDOS (see Fig.~\ref{bulkldosafm}) there are no bonding or
antibonding patterns for the two acceptor states, in agreement with the
results of the toy model. Rather, a spatially symmetric separation seems to
occur for some of the Mn pairs, in particular for larger distances along the
$\langle100\rangle$, $\langle211\rangle$ and $\langle111\rangle$ directions.
For these pairs the upper and lower acceptor acquire opposite and rather
definite spin characters in the easy direction. In this case spin up is
located on one site and spin down on the other, whereas for $\langle
110\rangle$ the two acceptors are of mixed spin character. \begin{figure}[ptb]
\resizebox{6.5cm}{!}{\includegraphics{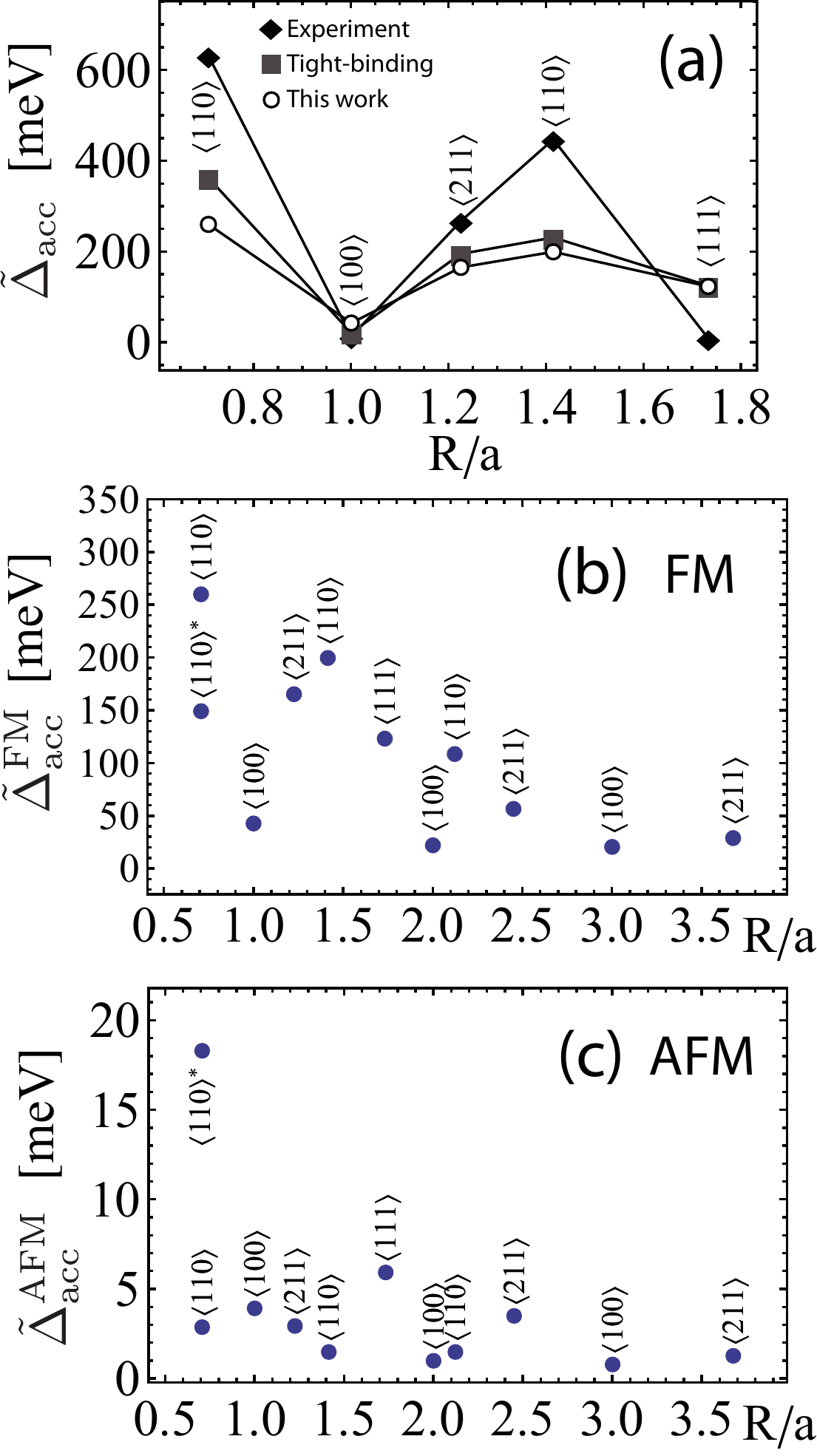}}\caption{Acceptor level
splittings as a function of pair orientation and distance. (a) shows a
comparative plot of current and theoretical and experimental results from Ref.
[\onlinecite{yazdani_nat06}]. In (b) and (c) the FM and AFM results for all
considered pairs are shown.}%
\label{splittings}%
\end{figure}

\subsubsection{Acceptor splitting vs. effective exchange constant $J$}

\label{acceptor_split_vs_J} In this Section we take a closer look at the
acceptor energy splitting $\Delta_{\mathrm{acc}}(\hat{\Omega})\equiv
\varepsilon_{2}-\varepsilon_{1}$, focusing in particular on its spherical
average $\tilde{\Delta}_{\mathrm{acc}}=\tilde{\varepsilon}_{2}-\tilde
{\varepsilon}_{1}$. We will also compute the effective exchange constant
$J(\hat{\Omega})$ and see if a relationship can be found between these two
important quantities.

Fig.~\ref{splittings} shows the average acceptor splitting $\tilde{\Delta
}_{\mathrm{acc}}$ in the FM and AFM configurations compared to the
experimental and theoretical values reported in
Ref.~[\onlinecite{yazdani_nat06,
yazdani_jap07}]. Our calculated splittings match the previous
calculations\cite{yazdani_nat06, yazdani_jap07}, and follow the trend of the
measured splittings at the (110) surface. Fig.~\ref{splittings} (b) reveals
that the splitting is largest for Mn pairs along $\langle110\rangle$
directions, and very small for $\langle100\rangle$ Mn pairs. The
$\langle211\rangle$ and $\langle111\rangle$ Mn pairs exhibit smaller
splittings than $\langle110\rangle$, which has more direct hopping paths.
Notice that there is some correlation between the splittings and the high
barriers in Fig.~\ref{highlow} (a); higher anisotropy barriers correspond to a
larger splitting.

Fig.~\ref{splittings} (a) shows that our tight-binding model as well as a
previous theory\cite{yazdani_nat06} based on a similar approach,
systematically underestimates the acceptor splitting measured experimentally.
This discrepancy could be due to inherent limited accuracy of the
tight-binding method. However, it is also quite possible that part of the
splitting seen experimentally is of Coulombic origin arising when electrons
tunnel into the two different acceptor levels. In Fig.~\ref{splittings} (c) we
see that the AFM splittings are very small compared to the FM ones, and are
typically of the order of a few meV. As already noted above, the only
exception is the $\langle110\rangle^{\ast}$ pair. Reducing the Coulomb
parameter on the common As neighbor dramatically lowers the acceptor energies
[see Fig. \ref{bulklevels} (b)], produces more extended wave functions (see
Fig. \ref{bulkldosafm}), and increases mixing with valence band states.

We can now consider the effective exchange energy defined as $J(\hat{\Omega
})=[E_{\mathrm{tot}}^{AFM}(\hat{\Omega})-E_{\mathrm{tot}}^{FM}(\hat{\Omega
})]/[2\cdot(5/2)^{2}]$, where the factors of $5/2$ from the Mn spin magnitudes
is absorbed in $J$. A positive (negative) $J$ implies that the Mn-Mn
interactions are FM (AFM). As a result of the spin-orbit interaction,
$J(\hat{\Omega})$ is an anisotropic quantity, that is, it depends on
$\hat{\Omega}$.

\begin{figure}[ptb]
\resizebox{6.5cm}{!}{\includegraphics{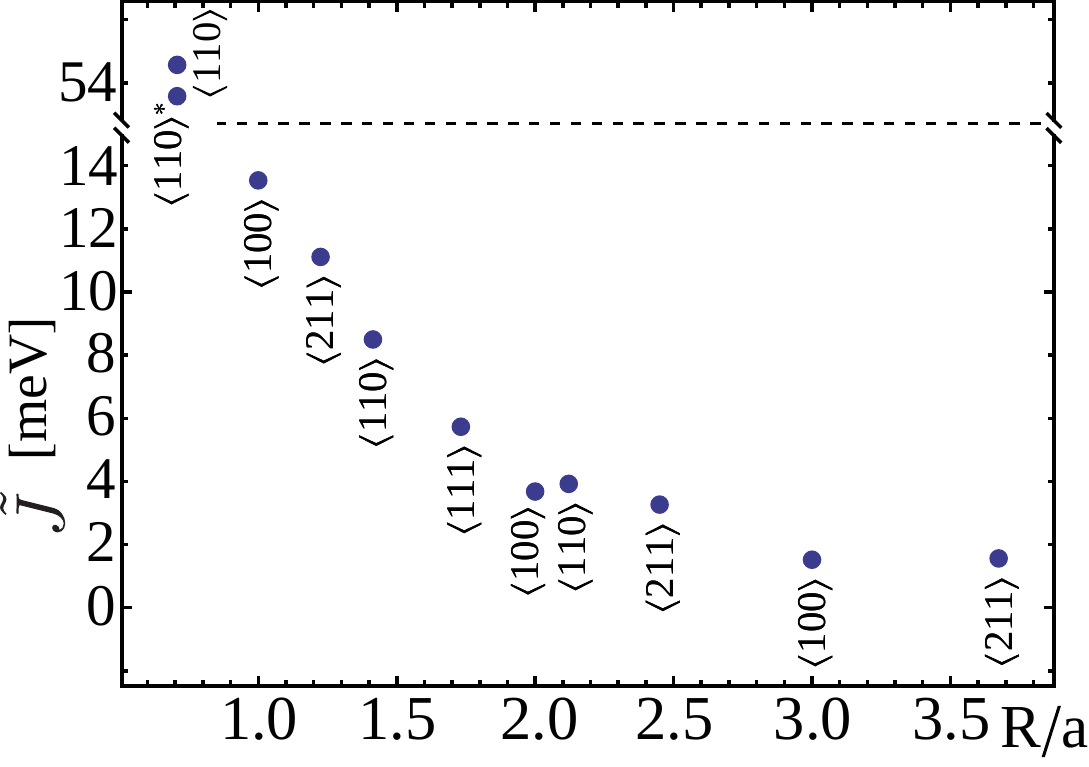}}\caption{Effective
exchange couplings. A ferromagnetic coupling is always favored as all
spherical averages $\tilde{J}>0$.}%
\label{jay}%
\end{figure}The effective $\tilde{J}$ as obtained by taking the spherical
averages is shown in Fig.~\ref{jay}, where it can be seen that the FM
configuration is always the most stable one, $\tilde{J}>0$. Only for
separations greater than two lattice constants, do the SO-induced fluctuations
in $J$ become comparable to the average value. Even then, ferromagnetism is
always favored. The maximal value of $J$ generally occurs when the Mn spins
are pointing along the pair axis.

The comparison between the acceptor splitting $\tilde{\Delta}_{\mathrm{acc}%
}^{\mathrm{FM}}$ for the FM state [Fig.~\ref{splittings} (b)] and the exchange
constant $\tilde{J}$ [Fig.~\ref{jay} (a)] shows that both quantities decay
rapidly with Mn separation. However, there are noticeable differences between
them, out of which two are most obvious. First of all, the relative value of
$\tilde{J}$ for the $\langle110\rangle_{d=0.71a}$ pair compared to all the
other values is much larger than the corresponding value of $\tilde{\Delta
}_{\mathrm{acc}}^{\mathrm{FM}}$. Secondly, $\tilde{\Delta}_{\mathrm{acc}%
}^{\mathrm{FM}}$ displays a much less monotonic decrease with Mn separation
than $\tilde{J}$. In particular, the large dip for the $\langle100\rangle
_{d=a,2a}$ pairs is totally absent in the plot for $\tilde{J}$. A similar dip
for the $\langle100\rangle_{d=2a}$ pair is hinted but much less pronounced in
the $\tilde{J}$ plot.

\begin{figure}[ptb]
\resizebox{6.5cm}{!}{\includegraphics{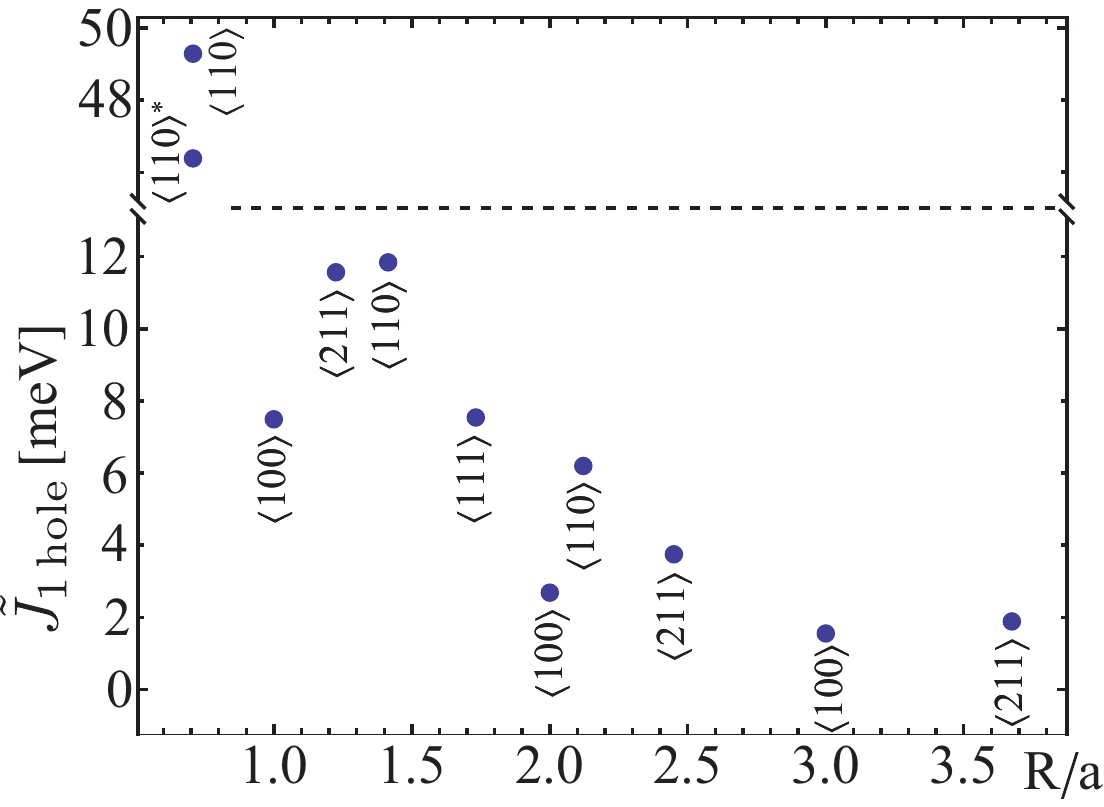}}\caption{Effective
exchange couplings for a system with two Mn and one hole. Adding one electron
to the system, gives an effective exchange that is related to the acceptor
splittings.}%
\label{jshift}%
\end{figure}

At this point it is useful to consider other quantities that can shed light on
the relationship between the effective exchange interaction and the acceptor
levels. We first consider the effective exchange constant for a Mn pair when
only one hole is present or, equivalently, when the lower acceptor is occupied
by an electron. The toy model results in Figs.~\ref{afm_pair} and
\ref{fm_pair_1} can help intuition and justify the rational of this choice:
when an extra electron is added to these electronic configurations, only the
topmost acceptor level is unoccupied in the FM state. In this case the
acceptor splitting should be directly related to the energy gain stabilizing a
FM state over the AFM state. The effect of occupying the lowest acceptor is
shown in Fig.~\ref{jshift}, where we plot $\tilde{J}_{\mathrm{1hole}}%
\equiv\lbrack\tilde{E}_{\mathrm{tot,1hole}}^{\mathrm{AFM}}-\tilde
{E}_{\mathrm{tot,1hole}}^{\mathrm{FM}}]/[2\cdot(5/2)^{2}]$, in which
$\tilde{E}_{\mathrm{tot,1hole}}^{\mathrm{AFM,FM}}$ is the spherical average of
the total ground state energy for a system with one Mn pair and only one hole,
namely with one extra electron added. We can indeed see that now the
dependence of $\tilde{J}_{\mathrm{1hole}}$ on the crystal orientation and
spacing of the Mn pair is qualitatively much more similar to the
$\Delta_{\mathrm{acc}}^{\mathrm{FM}}$ of Fig.~\ref{splittings} (b), displaying
the same large dips for the $\langle100\rangle$ pairs. \begin{figure}[ptb]
\resizebox{6.5cm}{!}{\includegraphics{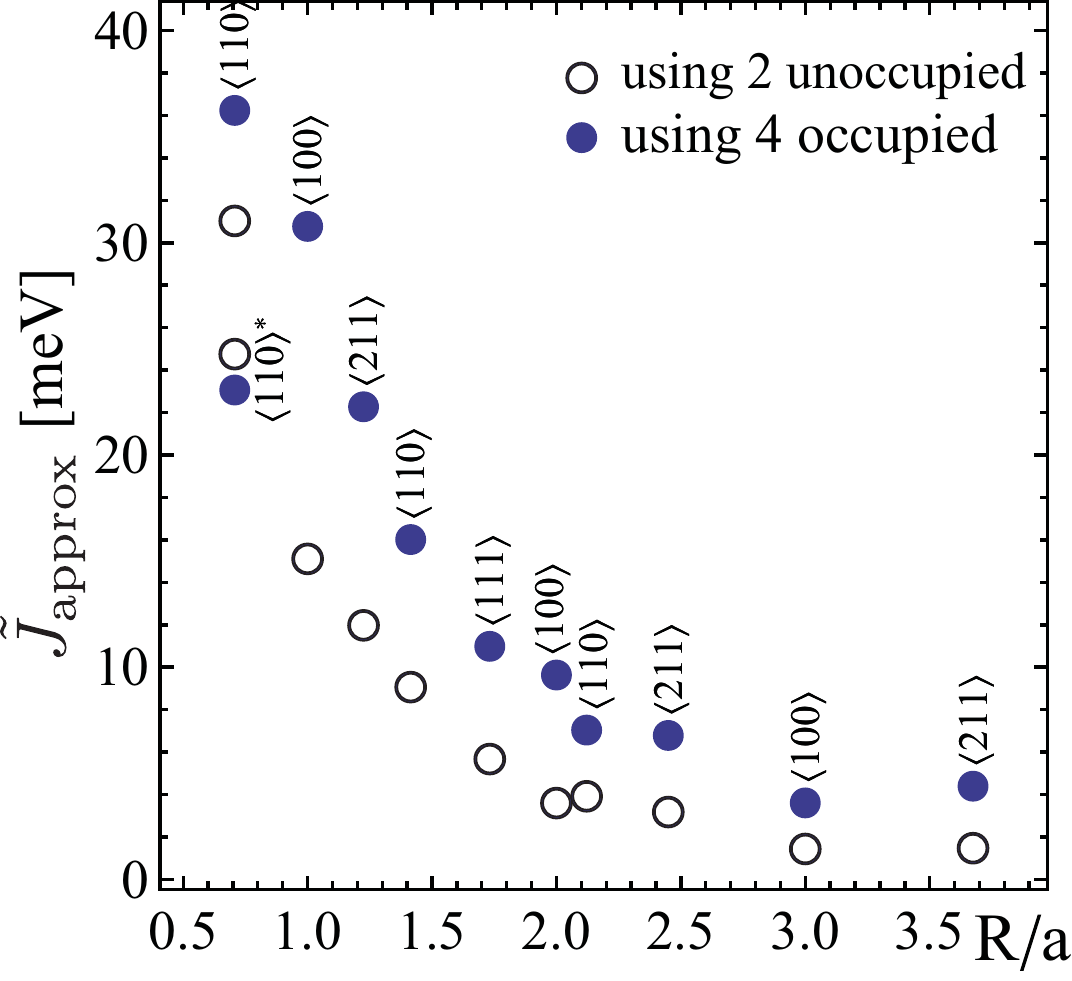}}\caption{The effective
exchange constant as estimated by the FM/AFM difference of the 4 occupied and
the 2 unoccupied acceptor levels. }%
\label{japprox}%
\end{figure}

In Fig.~\ref{japprox} we plot the spherical averages of the effective $J$ as
estimated by the FM/AFM difference of the sums of the 4 occupied impurity
levels and the 2 unoccupied levels. Comparing with the \textquotedblleft
exact\textquotedblright\ $\tilde{J}$ calculated by total energy difference of
the full system of electrons (Fig.~\ref{jay}), we see that both approximations
underestimate the value of $\tilde{J}$ for the most closely spaced
$\langle110\rangle$ Mn pair. For the rest of the points, using the four
highest occupied levels overestimates the value of $\tilde{J}$ approximately
by a factor of two. Surprisingly, taking the difference of the sums of the two
acceptors in the FM and the AFM configurations, gives a very good estimate of
the \textquotedblleft exact\textquotedblright\ $\tilde{J}$ for the rest of the
points in the plot of Fig,~\ref{jay} (a). In order to compute the effective
$\tilde{J}$ for the two acceptors, one must invert the sign, because 2
unoccupied levels are used.

\begin{figure}[ptb]
\resizebox{6.5cm}{!}{\includegraphics{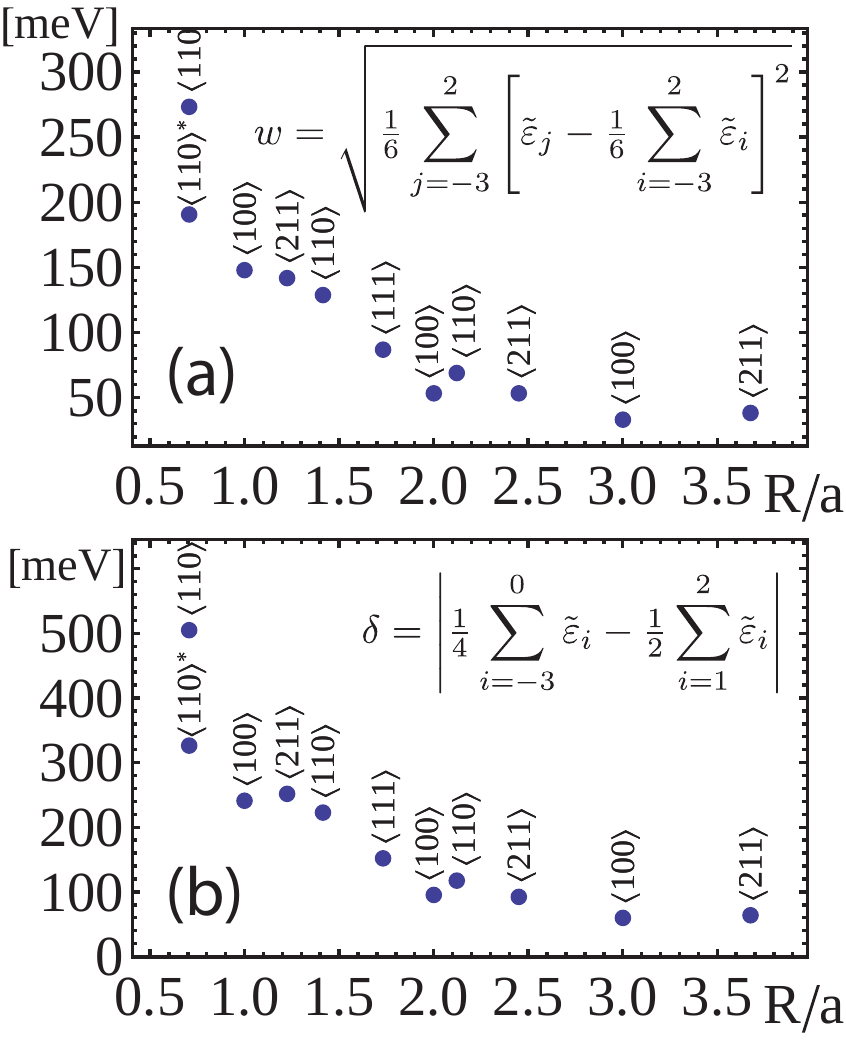}}\caption{Six impurity
level bandwidth (a), and splitting between the mean of the four occupied and
the mean of the two unoccupied impurity levels (b). }%
\label{altmeas}%
\end{figure}

In our discussion of the toy model, we have seen that the emergence of a
stable FM configuration for the Mn pair is brought about by the
\textquotedblleft widening\textquotedblright\ of a partially occupied cluster
of impurity levels caused by hopping. It is therefore instructive to compare
some measures of the \textquotedblleft bandwidth\textquotedblright\ of the six
impurity levels with $J$. Fig.~\ref{altmeas} (a) shows the FM effective
bandwidth of the six impurity levels,
\begin{equation}
w=\sqrt{\tfrac{1}{6}\sum_{j=-3}^{2}\left[  \tilde{\varepsilon}_{j}-\tfrac
{1}{6}\sum_{i=-3}^{2}\tilde{\varepsilon}_{i}\right]  ^{2}}, \label{effbw}%
\end{equation}
and Fig.~\ref{altmeas} (b) shows the splitting between the mean of the four
occupied impurity levels and the mean of the two unoccupied levels,%
\begin{equation}
\delta=\left\vert \tfrac{1}{4}\sum_{i=-3}^{0}\tilde{\varepsilon}_{i}-\tfrac
{1}{2}\sum_{i=1}^{2}\tilde{\varepsilon}_{i}\right\vert . \label{meansplit}%
\end{equation}
In the FM configuration these can both be seen as a measure of the effective
exchange interaction strength. A comparison of Figs.~\ref{altmeas} and
\ref{jay} reveals that they both do quite well in reproducing the correct
trend in $J$, with the exception of \ the most closely spaced $\langle
110\rangle$ pair. These observations support the validity of the simplified
arguments used in the toy model.

In conclusion, our analysis of the acceptor splitting versus effective
exchange constant shows that these two quantities are certainly related but
not in a direct quantitative way. In particular, the acceptor splitting is
much more anisotropic than $J$ as a function of the Mn pair orientation. Other
quantities such as $w$ and $\delta$ [see Eqs.~(\ref{effbw}) and
(\ref{meansplit})] involving all six impurity levels seems have a much better
correspondence with $J$. It is interesting to ask whether or not the four
occupied impurity levels are also accessible by STM. One might naively expect
that when the applied bias is reversed, electrons tunneling out of the
occupied impurity levels should appear as sharp features in the differential
conductance spectrum. The experiments of Ref.~[\onlinecite{yazdani_nat06}] do
not show any clear sign of these levels. However, a more recent
study\cite{gupta_pcm09} on individual Mn impurities in GaAs indicates that a
weak feature due to tunneling out of one the three $t_{2g}$ levels can indeed
be seen below the top of the valence band. Presumably this is due to a
hole-hole interaction effect: the final state has two electrons missing near
the Mn impurity. This feature is very hard to see, as it is masked by
electrons tunneling out of the valence band continuum.

\subsubsection{Comparison with ab initio estimates of the exchange coupling
$J$}

\label{abinitio_comparison} In this section we compare our results for the
effective exchange $J$ with those of previous studies. Since the pair exchange
interaction between impurities is a crucial quantity in the field of DMS,
there have been many theoretical studies of this quantity, mainly based on
first-principles calculations. Here we concentrate our attention on a few
issues that have emerged from our theoretical approach: (i) the relatively
large value for the $\langle110\rangle$ pair with shortest Mn spacing; (ii)
the non-monotonic decay of $J$ with Mn separation and its strongly anisotropic
character with respect to pair orientation; (iii) the comparison with acceptor
splitting, with particular reference to the discrepancies between $J$ and
$\Delta_{\mathrm{acc}}$ found in our model, e.g., for the $\langle100\rangle$
pairs. A few caveats are necessary before comparing our results with other
estimates of $J$ that have appeared in the literature. Firstly, most of the
published estimates are for much higher Mn concentrations, which strongly
affects the value of $J$ and its anisotropic properties. Secondly, all
first-principle calculations suffer from the well-known limitations of DFT
when applied to semiconductors. In particular, the estimates of $J$ depend on
the version of DFT used, e.g. LSDA vs. GGA. When the GGA+U approach is used to
account for electronic correlation effects, the value of the parameter $U$,
which has as strong influence on the results, can only be determined
indirectly by comparing with experiment.

In Ref.~[\onlinecite{bruno_prb04}] the electronic structure of (Ga,Mn)As was
calculated from first principles including disorder via the coherent-potential
approximation. The magnetic force theorem and one-electron Greens functions
then allow mapping onto an effective Heisenberg Hamiltonian. Their results for
the exchange interaction strength reveal a strong dependence on the doping
concentration $x$. In particular, the effective $J$ of the nearest
$\langle110\rangle$ pair is highly sensitive to doping and increases
dramatically with decreasing Mn concentration, ranging from $J=19$ meV for
$x=0.08$ to $J=55$ meV for $x=0.001$. The lowest concentration of 0.1\% Mn,
agrees very well with our estimate. Note that our calculations correspond to a
slightly smaller effective Mn concentration $x=0.0006$. In
Ref.~[\onlinecite{schilf_prb01}] the authors employed a self-consistent LSDA
atomic-spheres approximation\cite{okandersen_prb75} and their result of the
effective $J=55$ meV for the $\langle110\rangle_{d=0.71a}$ pair agrees
remarkably well with our values, despite the small supercell corresponding to
$x=0.009$. With spin-orbit interaction this value drops to 48 meV. In
Ref.~[\onlinecite{sarma_prl04}] the authors employ the GGA+U method within the
Projector Augmented Wave (PAW) \emph{ab initio} approach, treating $U$ as a
parameter. For $\langle110\rangle_{d=0.71}$ and $x=0.03$ they observe a
decreasing $J$ as a function of $U$, with $J=34$ meV for $U=0$ and $J=22$ meV
for $U=6$ eV. As the Mn concentration is reduced from $x=0.03$ to $x=0.008$ at
$U=0$, an increase in $J$ of approximately 10 meV is observed. A larger $U$
parameter causes the $d$ levels to emerge deeper in the valence band, the
acceptor wave function becomes more delocalized and $p$-$d$ exchange
decreases\cite{bruno_prb04b}. Photoemission experiments\cite{okabayashi_prb99}
indicate that the $t_{2g}$-states should be approximately 4 eV below the
valence band maximum, which corresponds to $U=3-4$ eV.
Ref.~[\onlinecite{sarma_prl04}] shows that the calculated $J$ depends on the
extension of the hole wave function as per chosen $U$.

We now turn to the $\langle100\rangle_{d=a}$ pair, which has a very low
experimental acceptor splitting (see Fig.~\ref{splittings}). In contrast to
our results (see Fig.~\ref{jay}), \emph{ab initio} calculations generally
predict\cite{bruno_prb04,olle_prb04,hilbert_prb05,sarma_prl04} a dip in the
curve of $J$ vs Mn separation occurring for this pair, with a lower value than
the two following points, $\langle211\rangle_{d=1.21a}$ and $\langle
110\rangle_{d=1.41a}$. Similar but smaller dips are also found for the other
$\langle100\rangle$ pairs.\footnote{One exception is again the LSDA
calculation of Ref.~[\onlinecite{schilf_prb01}], which finds $J=15$ meV, in
good agreement with our result, despite their small supercell corresponding to
$x=0.009$.} However, as mentioned above, all these calculations show a
significant dependence on impurity concentration. For example,
Ref.~[\onlinecite{bruno_prb04}] finds that the value of $J$ for the
$\langle100\rangle_{d=a}$ pair steadily increases from approximately $J=1$ meV
for $x=0.1$, to $J=7$ meV for $x=0.02$. So do the next two points for
$\langle211\rangle$ and $\langle110\rangle$ which end up at approximately 10
meV and 11 meV for $x=0.02$. When $U=0$ and $x=0.008$, the GGA + U approach
finds $J=6$ meV, well below the values for the next two pairs $\langle
211\rangle_{d=1.21a}$ and $\langle110\rangle_{d=1.41a}$, $J=14$ meV and $J=9$
meV respectively, which are also comparable to our results. Upon increasing
$U$ to the physically reasonable value of 6 eV, for both $\langle
211\rangle_{d=1.21a}$ and $\langle110\rangle_{d=1.41a}$ $J=7$ meV, while
$J=3.5$ meV for the $\langle100\rangle_{d=a}$ pair.

For even larger values of $U$ the difference in $J$ between the $\langle
100\rangle_{d=a}$ and $\langle211\rangle_{d=1.21a}$ pairs decreases further;
however, the $\langle100\rangle_{d=a}$ is still consistently lower. The GGA+U
values of $J$ for $\langle110\rangle_{d=1.41a}$ and $\langle211\rangle
_{d=1.21a}$, do not agree with the experimental acceptor splitting trend for
sound values of $U$. A similar trend for GGA calculations is reported in
Refs.~[\onlinecite{zunger_apl04,silva_prb04}]. The LDA
results\cite{sato_prb04} for 1\% Mn including disorder in the
coherent-potential approximation agree reasonably well with our results for
$\langle100\rangle_{d=a}$, $\langle211\rangle_{d=1.21a}$ and $\langle
110\rangle_{d=1.41a}$ with $J=8.2$, 9.1 and 7.9 meV. The $\langle
100\rangle_{d=a}$ is still consistently lower for larger distances and the
$\langle110\rangle_{d=0.71a}$ is relatively low at $J=18$ meV, indicating a
sensitivity of $J$ on the chosen \emph{ab initio} method.

In summary, with the caveats mentioned above, the general trend of the
\emph{ab initio} $J$ at reduced concentrations shows some qualitative
agreement with STM experimental acceptor splittings at shortest distances
$d=0.71a$, $d=a$. For larger Mn separations the agreement is not so good. As
corresponding theoretical estimates of the splittings are not reported in the
literature, it is hard to conclude whether or not a relationship between the
acceptor splittings and $J$ is present in the \emph{ab initio} calculations.
Concerning the comparison between the \emph{ab initio} values of $J$ and our
results, we can see that although the overall magnitudes agree reasonably
well, the qualitative trend differs since at short Mn separations we find a
monotonic decrease of $J$ with distance. \begin{figure}[ptb]
\resizebox{6.5cm}{!}{\includegraphics{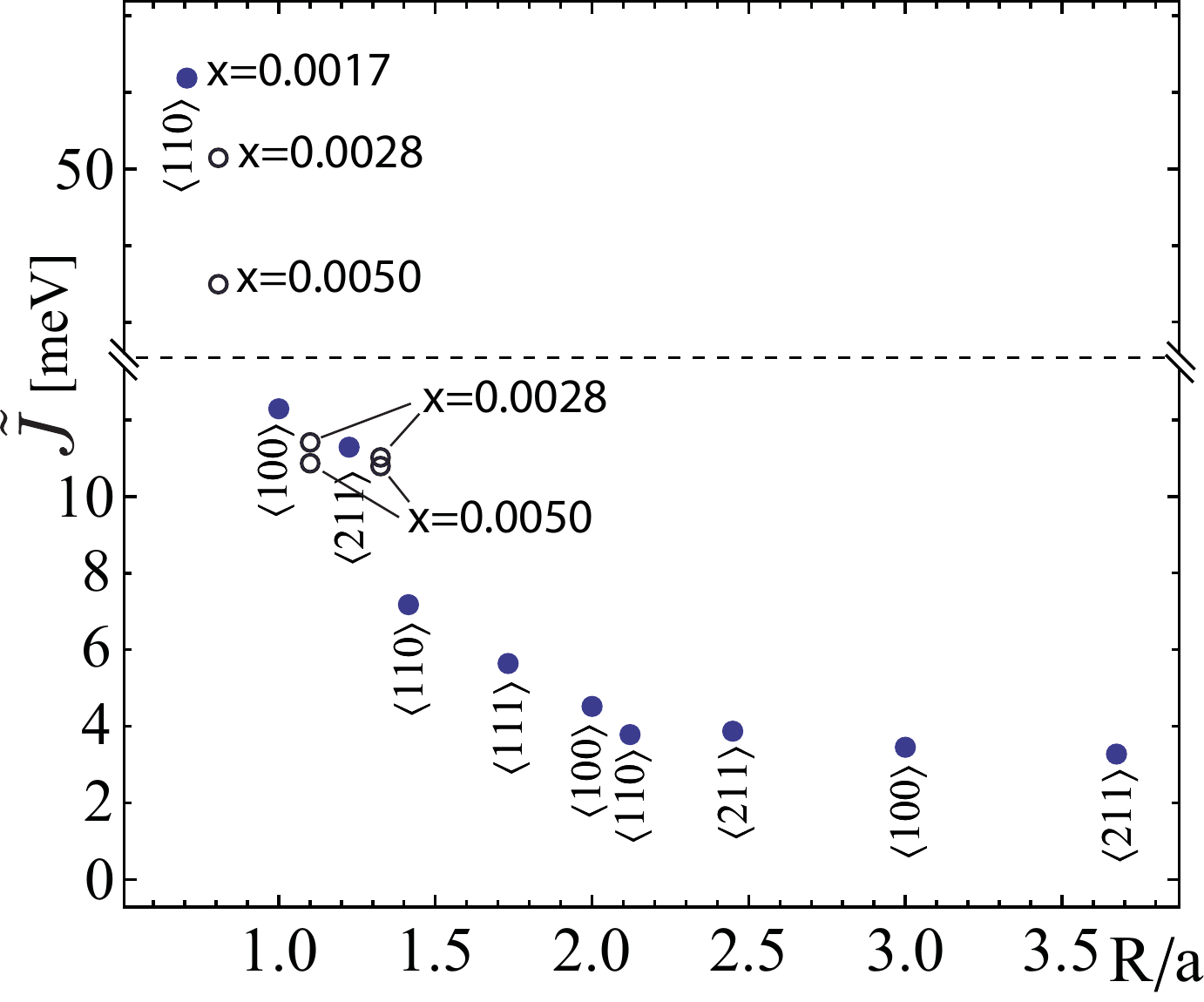}}\caption{The effect on
$\tilde{J}$ when increasing the Mn concentration from $x=0.0006$ (shown in
Fig.~\ref{jay}) to $x=0.0017$ (filled circles). The empty circles show how
$\tilde{J}$ changes when increases $x$ further for selected points.}%
\label{hjav}%
\end{figure}

Given the strong dependence of $J$ on Mn concentration and the fact that the
reported \emph{ab initio} results are obtained for larger concentrations, it
is interesting to investigate how our value of $J$ changes when we increase
the Mn concentration. In Fig.~\ref{hjav} the effect of decreasing the
supercell size such that the Mn concentration changes from $x=0.0006$ to
$x=0.0017$ is shown. The $\langle100\rangle_{d=a}$ $J$ is now closer to the
$\langle211\rangle_{d=1.21a}$ value. Increasing the concentration further
causes the two pairs to obtain equal $J$, but the $\langle100\rangle_{d=a}$
value never drops below the $\langle211\rangle_{d=1.21a}$ value, indicating
that there is a fundamental difference with respect to \emph{ab initio}.

Generally, \emph{ab initio} predicts a qualitative trend that agrees better
with our effective $J$ obtained by occupying the lower acceptor (see
Fig.~\ref{jshift}). The $\langle100\rangle_{d=a}$ highest occupied level and
lower acceptor are quasi-degenerate, with a gap that varies between 14-46 meV.
In \emph{ab initio} methods, a common technique used to speed up evaluation of
k-point sums, is to introduce a fractional occupation of the unoccupied levels
controlled by an occupation smearing parameter. At the end of the calculation,
the limit of zero smearing is taken to calculate the total ground state
energy. It is still possible that the end result depends on the choice of
smearing parameter. If this is the case, the effect can cause the $J$ for the
quasi-degenerate $\langle100\rangle_{d=a}$ to decrease.

\subsection{Mn-Mn interactions in the (110) GaAs surface}

\label{surfresults} So far we have studied Mn pairs embedded in bulk GaAs. In
this last Section we consider the experimentally more relevant situation where
the two Mn atom substitute two Ga atoms on the $(110)$ surface. We consider
again a 3200-atom supercell, but this time we apply periodic only in the two
directions in the plane of the $(110)$ surface. This corresponds to a
$38\times38$ A$^{2}$ surface and a supercell cluster that has 20 atomic layers
along the surface normal separating the two (110) surfaces, such that
surface-surface interactions are negligible. The loss of coordination and
hybridization with the surface states will cause the impurity levels to appear
very deep in the gap. The depth is subject to experimental uncertainty due to
a band-bending\cite{garleff_prl09} effect when imaging a semi-conductor
surface with a metal tip. In our model, the depth is very sensitive to the
off-site Coulomb correction, $V_{\mathrm{off}}$, and we use this parameter to
reproduce the acceptor level of a single Mn on the (110) surface at the
experimentally observed position\cite{yazdani_nat06} at 850 meV. The parameter
$V_{\mathrm{off}}$ has therefore been reduced from 2.4 eV to 1.57 eV.

We proceed again to study the magnetic anisotropy energy of the systems, the
properties of the mid-gap acceptor states, the effective exchange interaction
and its connection with the acceptor splitting.

\subsubsection{Magnetic anisotropy energy}

\label{surface_anysotropy}

\begin{figure}[ptb]
\resizebox{7.5cm}{!}{\includegraphics{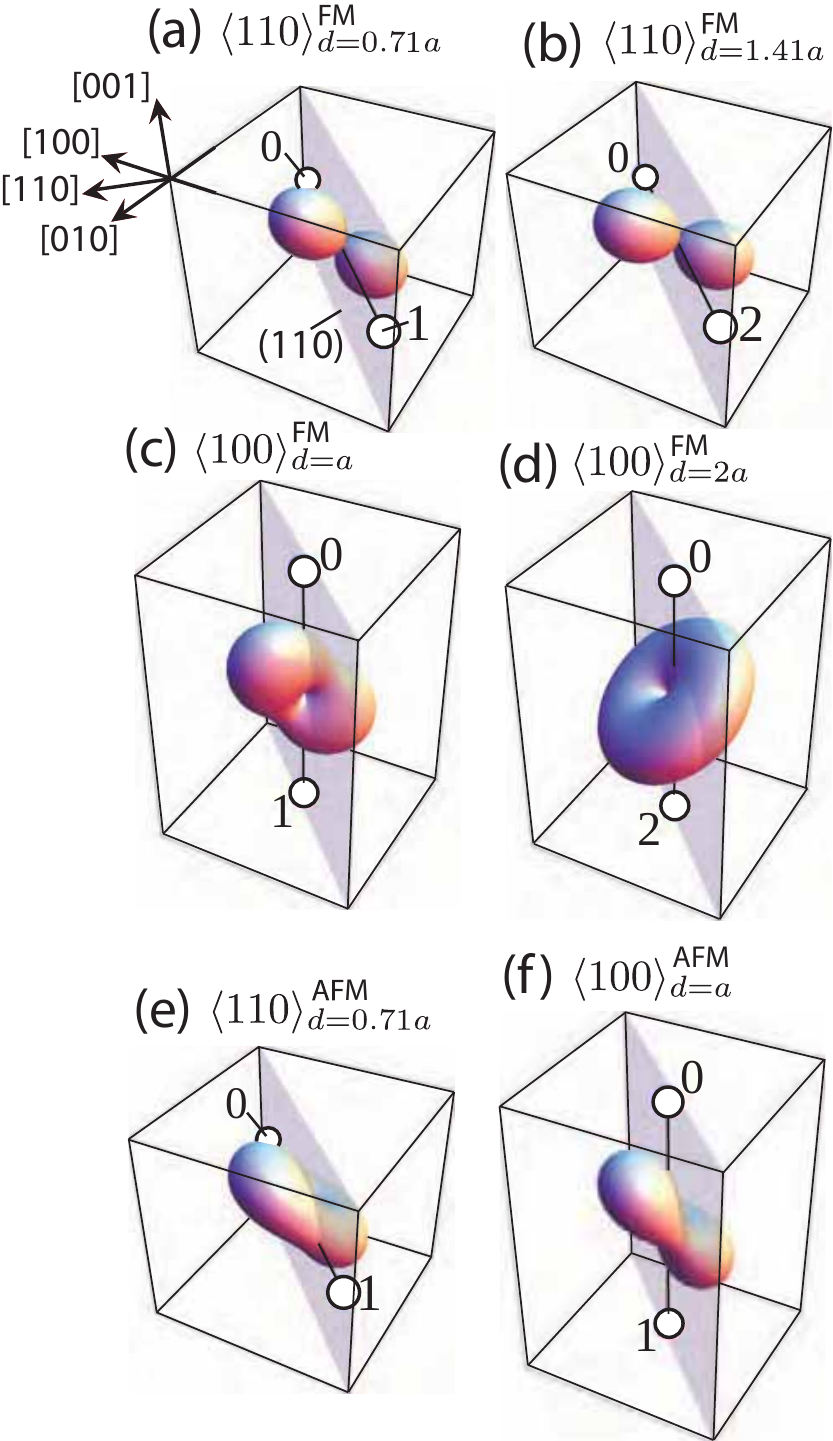}}\caption{(Color online)
The magnetic anisotropy landscapes for Mn pairs in the (110) surface. The
transparent plane indicates the (110) surface. (a)-(d) show the distinct types
of landscapes in the FM configurations. Only along the $\langle110\rangle
_{d=0.71a,1.41a}$ in (a) and (b), and $\langle100\rangle_{d=a}$ in (c), do the
landscapes differ from the result of weakly interacting pairs in (d), which is
just the sum of two isolated Mn anisotropies. Similarly, in the AFM
configurations, only $\langle110\rangle_{d=0.71a}$ (e) and $\langle
100\rangle_{d=a}$ (f) deviate qualitatively from the landscape (d). }%
\label{smae}%
\end{figure}\begin{figure}[ptb]
\resizebox{6.5cm}{!}{\includegraphics{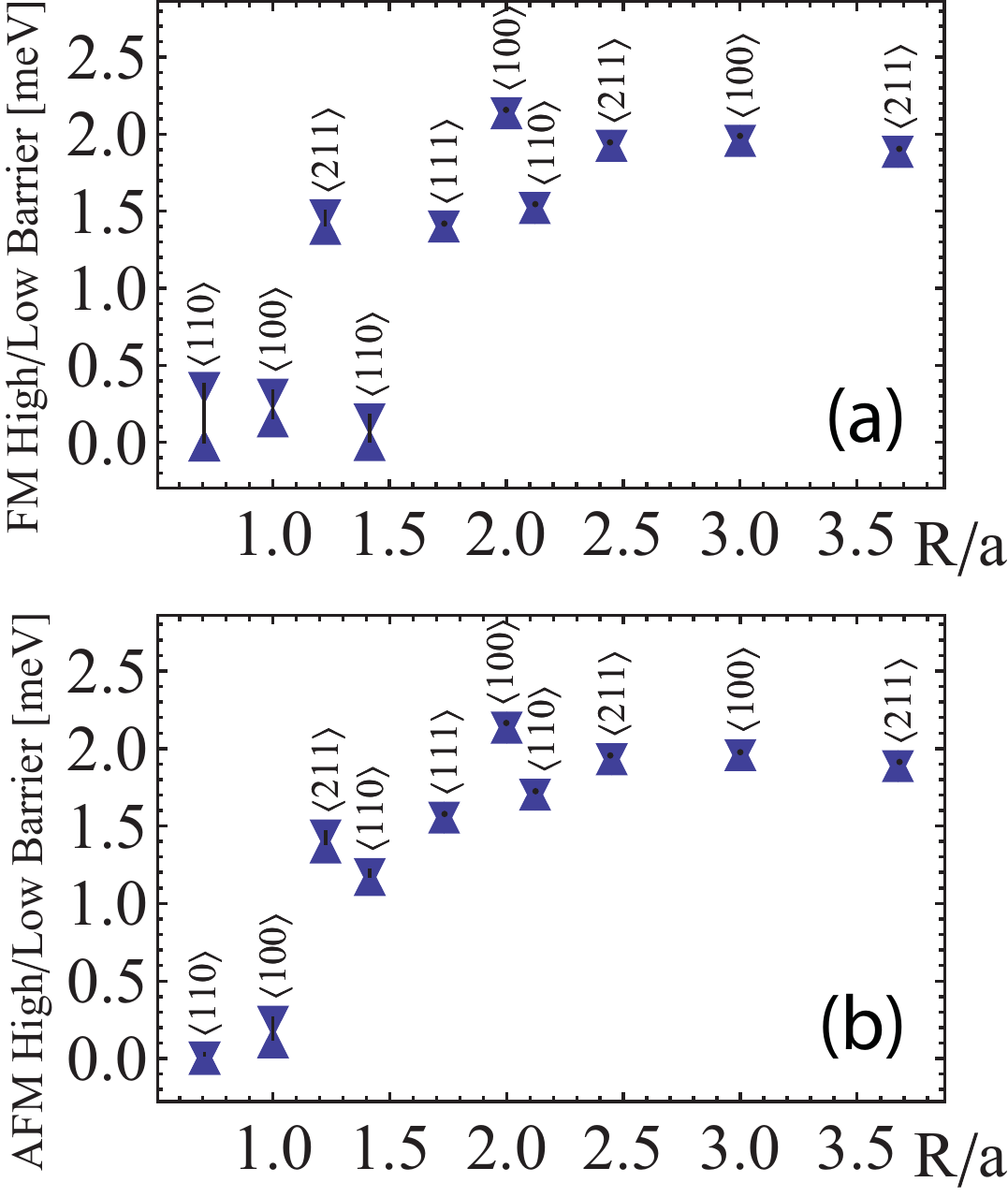}}\caption{The high and low
barriers for the FM (a) and the AFM (b) configurations for pairs at the (110)
surface. The anisotropy for the closest pairs exhibit a reduced anisotropy
energy for separations greater than two lattice constants, all pairs have a
characteristic barrier of around 2 meV, indicative of very weakly interacting
Mn. }%
\label{safmfmvar}%
\end{figure}Fig.~\ref{smae} shows the qualitatively distinct anisotropy
landscapes for Mn pairs on the surface and Fig.~\ref{safmfmvar} the magnitude
of the barriers. The surface geometry completely dominates the anisotropy and
for most separations the anisotropy landscape is qualitatively the same as for
a single Mn in the surface\cite{scm_MnGaAs_paper1_prb09} [exemplified in
Fig.~\ref{smae} (d)]. This type of anisotropy landscape occurring for
separations larger than $\approx1.5$ lattice constants is an indication that
the acceptor state hybridization is so weak that the resulting anisotropies
are essentially the sum of two independent Mn atom anisotropies. Only for the
closest pairs do different anisotropies appear. In the FM configurations only
$\langle110\rangle_{d=0.71a}$, $\langle110\rangle_{d=1.41a}$ and
$\langle100\rangle_{d=a}$ [Fig.~\ref{smae} (a)-(c)] show qualitatively
different landscapes - tilted quasi-easy planes with the hard direction
approximately along [111]. Similarly for the AFM configurations, only
$\langle110\rangle_{d=0.71a}$ and $\langle100\rangle_{d=a}$ [Fig.~\ref{smae}
(e)-(f)] differ from the single Mn type landscape in Fig.~\ref{smae} (d). In
$\langle100\rangle_{d=a}$ the easy axis and the low barrier are interchanged
between the AFM and FM, but the hard direction remains along the [111].

Fig.~\ref{safmfmvar} shows that anisotropy energies are very small, typically
one order of magnitude smaller than for the fully periodic systems. In both FM
and AFM configurations the interactions between Mn for distances below two
lattice constants, tend to reduce the anisotropy heavily. At larger distances,
where all Mn pairs produce the [111] easy axis type landscape in
Fig.~\ref{smae} (d), the single barriers are around 2 meV. This limit of
weakly interacting Mn is also reflected in the difference in barriers (see
\ref{safmfmvar}) between the FM and AFM, which becomes negligible for weakly
hybridized pairs above $1.5a$. As we will show, total energy differences are
also small, causing a very small effective exchange. \begin{figure}[ptb]
\resizebox{5.5cm}{!}{\includegraphics{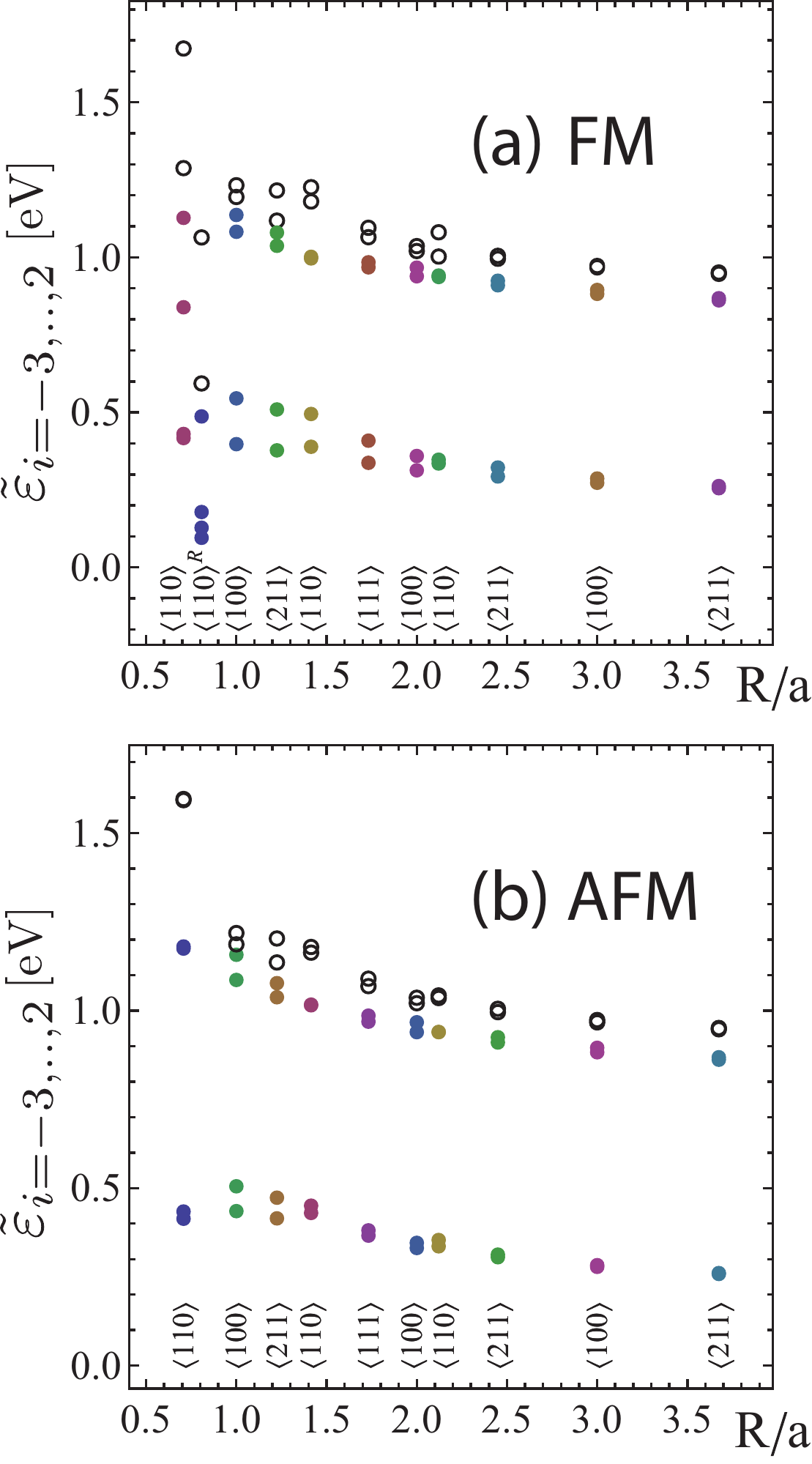}}\caption{The four highest
occupied (filled circles) and the two acceptor levels (empty circles) in the
FM (a) and AFM (b) configuration for Mn pairs in the (110) surface. Loss of
coordination and hybridization with surface states cause impurity levels deep
in the gap.}%
\label{surflevels}%
\end{figure}

\subsubsection{Character of acceptor levels for Mn pairs in a (110) surface}

Fig.~\ref{surflevels} shows the group of six impurity levels for the Mn pairs
in the surface. We see that for both FM and AFM configurations, two of the
occupied levels are split around 0.5 eV below the others. For the closest
$\langle110\rangle$ pair the acceptors are so high in energy that the upper
acceptor has in fact crossed the first conduction band state. This is
associated with a very large splitting of the acceptors. In $\langle
110\rangle^{R}$ the off-site Coulomb has been reduced even further (to 0.5 eV)
in order to also make the upper acceptor appear in the gap. The overall
spectrum for the AFM and FM configuration looks similar, indicating that
interactions are very weak and the itinerant spin wave functions are very
localized at the surface. In the STM experiment\cite{yazdani_nat06} Zn dopants
can give rise to resonant tunneling between conduction states and the acceptor
level, thought to be responsible for the negative dip seen in the curve of the
tunneling conductance vs. bias voltage.\cite{loth_prl06} The coupling to
conduction-like states can lead to a more extended, bulk-like acceptor wave
function seen at the (110) surface. As a first approximation our treatment of
the surface should reveal some of the relevant properties. Our results
indicate that the interactions in the surface are much weaker than in bulk,
and magnetic anisotropies are one order of magnitude smaller. This is related
to acceptor wave functions that are much more localized at the surface.
\begin{figure}[ptb]
\resizebox{6.2cm}{!}{\includegraphics{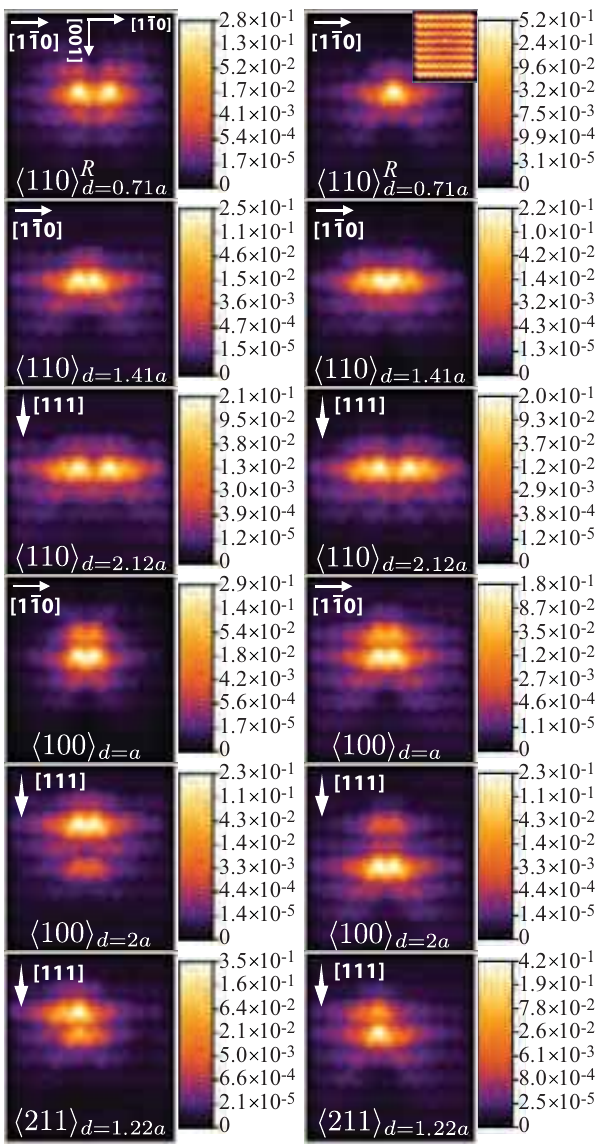}}\caption{(Color online)
LDOS for FM aligned Mn pairs on the (110) surface. The inset in $\langle
110\rangle^{R}$ shows the level above the low acceptor when the upper acceptor
has crossed into the conduction band. The images are for the Mn spin in the
easy direction, indicated in the top left corner of each panel (note that
[111] is coming out of the (110) surface).}%
\label{sldosfm}%
\end{figure}\begin{figure}[ptb]
\resizebox{6.2cm}{!}{\includegraphics{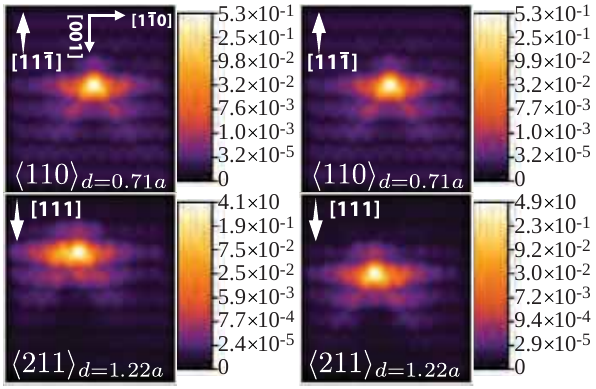}}\caption{(Color online)
Examples of LDOS for AFM aligned pairs Mn in the (110) surface.}%
\label{sldosafm}%
\end{figure}

So far there are relatively few papers that attempt to simulate STM images of
Mn in the (110) surface.\cite{mikkelsen_prb04,stroppa_prb07} The acceptor LDOS
resulting from our tight-binding model for a few representative pairs are
shown in Fig.~\ref{sldosfm} for the FM and Fig.~\ref{sldosafm} for the AFM
configurations. The spectral weight in the core regions of the Mn are much
higher than in bulk. As an example, the upper acceptor for the $\langle
110\rangle_{0.71a}^{R}$ has around 50\% spectral weight on the As in-between
the Mn, to be compared with 17\% in bulk, which is still high relative other
bulk configurations. Wave functions for other surface pairs have a much higher
maximum spectral weight than in bulk, with values in the typical range
20\%-30\%. Overall, the spectral weight maximum for a given pair in bulk is
approximately an order of magnitude smaller than the corresponding pair at the
surface, with bulk wave functions that are much more spread out in the lattice.

As shown in the first two rows of Fig.~\ref{sldosfm}, some bonding/antibonding
characteristics in the acceptor wave functions can be seen for the closest
pairs $\langle110\rangle_{d=0.71a,1.41a}$ in the FM configuration, which
exhibit magnetic quasi-easy planes. Both acceptor wave functions for the
$\langle110\rangle_{d=2.12a}$ Mn pair exhibit a hint of antibonding character.
Note that this pair is characterized by the magnetic anisotropy landscape of a
single Mn impurity.

For $\langle100\rangle_{d=a}$ each acceptor level is occupying both Mn sites,
but a separation occurs for $\langle100\rangle_{d=2a,d=3a}$ such that each
acceptor has more pronounced LDOS on one of the sites. The same association of
acceptors spatially bound to one site, appears for the other directions - a
pattern indicative of weak hybridization. Finally the $\langle211\rangle
_{d=1.22a}$ has some hint of bonding/antibonding pattern, but not as
pronounced as for bulk.

In the LDOS for the AFM configurations (see Fig.~\ref{sldosafm}), the
$\langle110\rangle$ pairs show no spatial separation between the acceptors and
no bonding/antibonding behavior. The $\langle211\rangle_{d=1.22a}$ has upper
and lower acceptor wave functions that are localized to the separate Mn sites,
whereas the rest of the pairs are similar to the bulk counterparts (see Fig.
\ref{bulkldosafm}), but with a more localized LDOS signature.
\begin{figure}[ptb]
\resizebox{6.0cm}{!}{\includegraphics{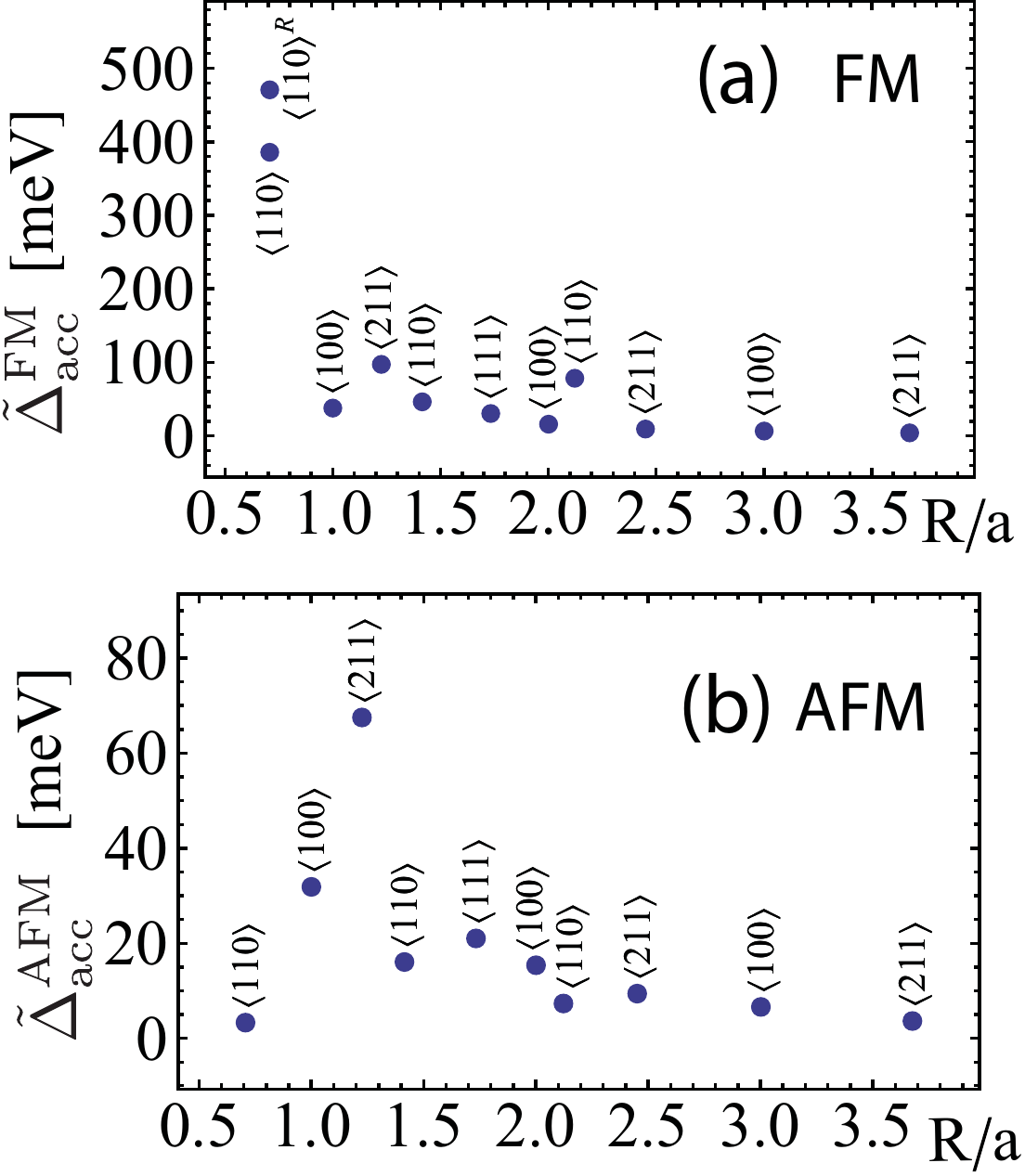}}\caption{The splittings of
the acceptor levels for Mn pairs in the (110) surface. }%
\label{surfsplit}%
\end{figure}\begin{figure}[ptb]
\resizebox{6.0cm}{!}{\includegraphics{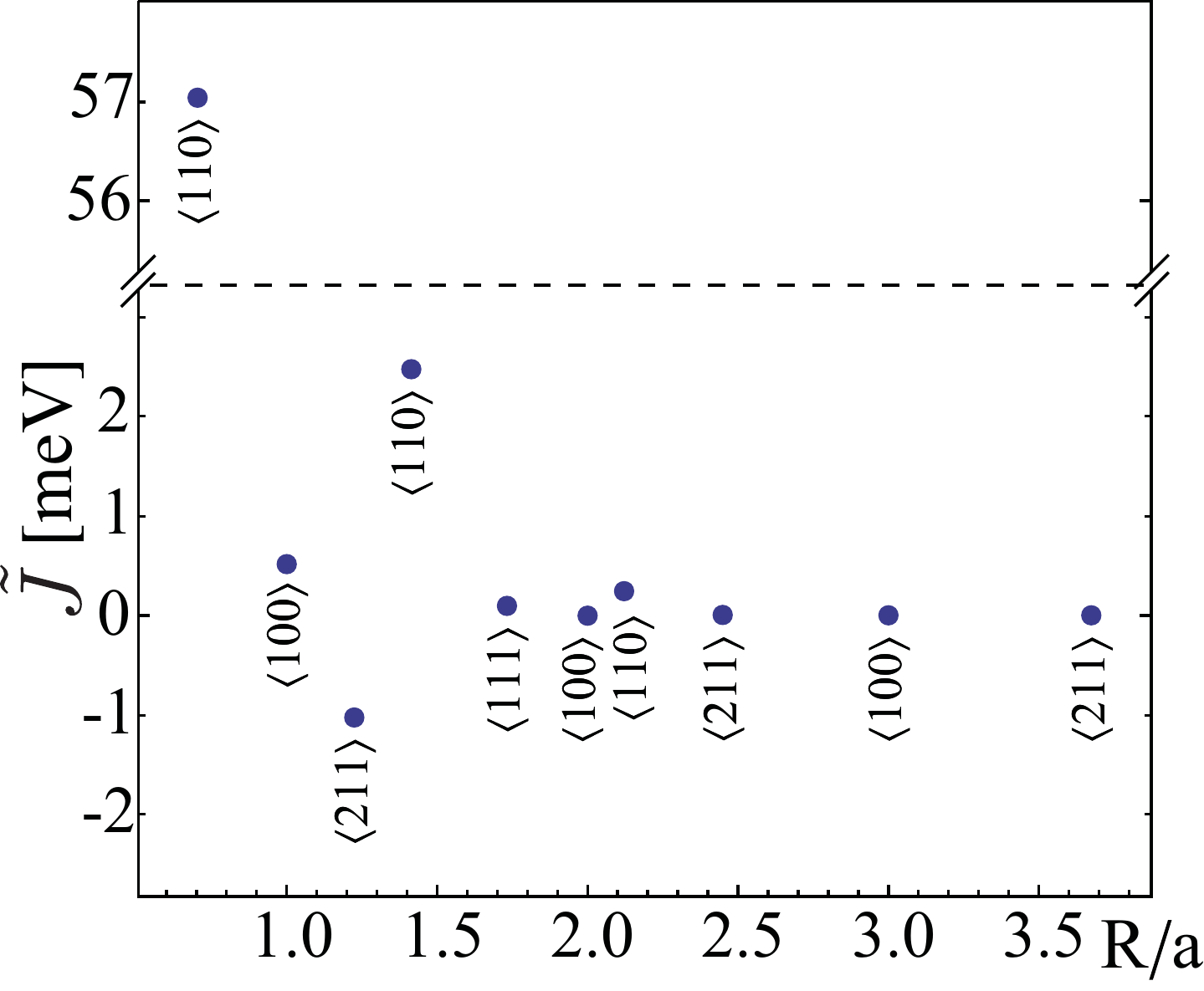}}\caption{The effective
exchange constant for Mn pairs in the (110) surface. The effective $\tilde{J}$
falls off much faster at the surface, due to the highly localized acceptor
wave functions. The $\langle211\rangle_{d=1.22a}$ is showing antiferromagnetic
behavior.}%
\label{sjav}%
\end{figure}

\subsubsection{Acceptor splitting vs. effective exchange constant $J$}

\label{surface_acceptor_split_vs_J}Finally, we have calculated the effective
exchange constant for the Mn pairs in the surface, which we compare with the
calculated acceptor splittings. The FM splittings of the acceptor levels is
shown in Fig.~\ref{surfsplit} (a). The very large splitting of 600 meV for the
closest Mn pair seen in experiment (see Fig.~\ref{splittings}) is now
comparable to the calculated one of around 500 meV, when both acceptors are in
the gap. The calculated splitting is small for the nearest $\langle100\rangle
$. It then increases to around 100 meV for the $\langle211\rangle$, but after
that splittings are very small and less than 50 meV, with the exception of the
farthest $\langle110\rangle$. In the AFM configurations, splittings are much
larger at the surface compared to bulk. In particular, the closest
$\langle211\rangle$ has a large splitting of around 70 meV. This is expected,
since a finite splitting in the AFM configuration is possible when inversion
symmetry is broken, an effect that is clearly enhanced by the presence of the surface.

The FM surface splittings do not match the long range experimental trend as
well as the bulk calculations. In our calculations, we have accounted for the
buckling of the GaAs (110) surface, but not the detailed relaxation of Mn
environment. It is also not certain that the effective $p$-$d$ exchange model
holds so well for the surface. Not only the long range and short range Coulomb
may differ at the surface, but the exchange splitting is also subject to
change. This is because the $p$-$d$ hybridization scheme in
Fig.~\ref{cartoon2} will be altered by the surface states.

The surface $\tilde{J}$ are shown in Fig.~\ref{sjav}. For the nearest
$\langle110\rangle$ pair, $\tilde{J}$ is strongly ferromagnetic and has a
value similar to the corresponding pair in bulk. The $\langle100\rangle$ pair
has a very low $\tilde{J}$ of around half an meV and could easily become
antiferromagnetic by a change in Coulomb parameters. The SO-induced anisotropy
of $J$ on the unit sphere of magnetic moment directions is very weak, just
fractions of an meV, due to the low anisotropy energies at the surface.
Interestingly the $\langle211\rangle$ pair, which has a relatively large
acceptor splitting of several tens meV in the AFM configuration (see
Fig.~\ref{surfsplit}), is antiferromagnetic. As mentioned above, this is an
effect entirely due to the surface, where a significant acceptor splitting can
exist even for antiferromagnetically aligned Mn spins and generate an energy
gain that can compete with the FM configuration. The rest of the Mn pairs
shown in Fig.~\ref{sjav} have a very small $\tilde{J}$, indicating that the Mn
local moments interact very weakly due to the high degree of localization of
the acceptor wave functions at the surface. The oscillatory behavior is
indicative of Friedel oscillations with a much shorter period than in bulk,
which is connected to the strong localization of the acceptor wave functions
at the surface. Finally, it is interesting to note that the magnitude of
$\tilde{J}$ (disregarding the sign), follows the trend of the experimentally
observed splittings, despite the fact that the calculated ones do not clearly
do so. Disregarding the single point $\langle110\rangle_{d=1.41a}$ in
Fig.~\ref{surfsplit} (a), the experimental trend but with smaller magnitudes
is reproduced. It is also interesting to note the large AFM splitting for
$\langle211\rangle_{d=1.22a}$ in Fig.~\ref{surfsplit} (a), which makes it
difficult to distinguish from the FM configuration. There are however
noticeable differences between the $\langle211\rangle_{d=1.22a}$ FM and AFM
LDOS (see Figs.~\ref{sldosfm} and \ref{sldosafm}). The FM acceptors each have
spectral weight on both Mn sites, but the AFM acceptors are each associated
with one site. Given the large splitting between the AFM acceptors, this
observation could perhaps be tested experimentally. Overall, it would appear
that the experimental results for the (110) surface are intermediary to our
surface and bulk calculations.

\section{Conclusions and Outlook}

\label{conclusions}

In this paper we have investigated substitutional Mn pairs in GaAs, using a
microscopic tight-binding model solved for large GaAs clusters. The model
accounts realistically for spin-orbit interactions and includes the
kinetic-exchange $p$-$d$ coupling between the Mn local magnetic moment and the
spin of valence band electrons. One goal of this work was to assess whether or
not this model can reproduce the main features of coupled acceptor states
observed in recent STM experiments,\cite{yazdani_nat06,yazdani_jap07} and in
this way further elucidate the nature of the effective exchange interaction
between the two Mn magnetic moments. When the two Mn atoms are located in the
bulk, we find that the ground state of the system is generally one in which
the two Mn magnetic moments are coupled ferromagnetically and aligned along
the direction connecting the Mn atoms. In the ferromagnetic configuration, the
two topmost acceptor states are split, and their energy separation depends
strongly on the Mn pair orientation with respect to the GaAs crystal structure
and decreases with Mn separation. In particular, the largest splitting
($\sim300$ meV) is found for the Mn pair oriented along the $\langle
110\rangle$ direction. For separations larger than three lattice constants,
the splitting is below 20 meV. The splitting is the result of hybridization of
degenerate acceptor states with identical spin character, giving rise to
bonding and antibonding states. Our calculated acceptor splittings are
typically a factor of two smaller than the experimental values. Coulomb
correlation effects not included in our tight-binding approach could partly be
responsible for the larger splittings. Analysis of the LDOS for the two
acceptor states shows that these are mainly concentrated in the surroundings
of the Mn atoms; for intermediate separations, when the splitting is sizable,
the higher state and lower state are of the \textit{bonding} and anti-bonding
type respectively. These results are all consistent with
experiment,\cite{yazdani_nat06,yazdani_jap07} indicating that the TB model
correctly describes the electronic properties of the coupled acceptor states
in GaAs associated with the magnetic impurities.

An important question that we have addressed here is whether or not the
acceptor splitting, which is accessible by STM measurements, is directly
related to the effective exchange coupling between the Mn local moments. The
present work demonstrates that, at least within the model considered here,
this relationship is not very sharp: although both quantities decrease with
separation and are typically anisotropic, the acceptor splitting displays a
non-monotonic behavior for a few pair orientations which is not found in the
exchange coupling. In our model, the dependence of the latter on Mn separation
and pair orientation is better represented by the overall \textit{bandwidth}%
\cite{endnote50} of the six impurity levels present in the gap (see
Fig.~\ref{altmeas}). In particular, Our calculations demonstrate that exchange
interactions between two Mn ions in (Ga,Mn)As are closely related to the
splitting between the two acceptor levels that are occupied by holes and the
four lower energy acceptor levels that are occupied by electrons. This
splitting is not readily measured by STM experiments which are strongly
influenced by Coulomb interaction energies when the number of holes bound to
the acceptor complex is increased above two by removing electrons from the
system. It is possible that infrared spectroscopy of transitions between
iso-charge levels of the acceptor complex could be more effective.

These results support the idea that the stability and strength of the
ferromagnetic coupling involves the hybridization of all impurity levels and
not only the two topmost ones occupied by holes. Our estimate of the exchange
coupling is in qualitative agreement with results obtained from \emph{ab
initio} calculations, with a few noticeable discrepancies analyzed in detailed
in Sec.~\ref{abinitio_comparison}, which are most likely due to the
limitations of both methods and deserve further investigation.

When the two Mn atoms are inserted in the $(110)$ surface as in experiment, we
typically find much deeper acceptor states with strongly localized wave
functions around the impurities. Weaker hybridization causes the acceptor
splitting to decay more rapidly with separation, although in a manner
qualitatively similar to bulk. With the exception of the most closely spaced
Mn pair, the surface acceptor splittings seem to agree less well with
experiment than our bulk results. The exchange constant is likewise very small
except at the shortest separations and can even change sign, favoring an
antiferromagnetic coupling. These results are a clear indication that our
semiphenomenological microscopic model, derived mainly from bulk (Mn,Ga)As
properties, is not able to quantitatively capture the complexity of the
surface states. This is an area that could benefit from additional work.

The model investigated here makes predictions for the spin-orbit induced
magnetic anisotropy for pairs of Mn atoms in GaAs. Some of these predictions
could be tested in STM experiments by applying an external field. When the Mn
pair is in a bulk environment, there is a uniaxial magnetic anisotropy along
the axis of the pair with anisotropy barriers of 10-20 meV. When the pair is
in a (110) surface, the behavior of the anisotropy energy landscape is for
most pairs similar in character and magnitude to that of a single Mn atom in
the (110) surface.\cite{scm_MnGaAs_paper1_prb09}.

In the present approach the local moment degrees of freedom have been treated
as classical variables. The study of the quantum spin dynamics of two
interacting Mn in GaAs is an important and interesting subject both
theoretically and in view of future experiments. Work in this direction, based
on Berry-phase quantization of the semi-classical local moments considered
here, is underway.

\section{Acknowledgments}

This work was supported by the Welch Foundation and the National Science
Foundation under grant DMR-0606489, the Faculty of Natural Sciences at Kalmar
University, the Swedish Research Council under Grant No: 621-2004-4439, and by
the Office of Naval Research under grant N00014-02-1-0813. We would like to
thank J. Gupta, A.~Yazdani, P.~M.~Koenraad, J.~K.~Garleff, A.~P.~Wijnheijmer
and C.~F.~Hirjibehedin for useful discussions.


\end{document}